\documentclass[epsf,useAMS,usenatbib]{mn2e}
\usepackage[usenames,dvipsnames]{color}
\usepackage{float}  
\usepackage[pdftex]{graphicx}

\usepackage{titlesec}
\def\mnras{MNRAS}
\def\apj{ApJ}
\def\aap{A\&A}
\def\araa{ARA\&A}
\def\na{New A}
\newcommand{\apjs}{ApJS}		
\newcommand{\icarus}{Icarus}		
\newcommand{\apss}{Ap\&SS}		
\newcommand{\Msun}{\hbox{$\hbox{M}_\odot\;$}}
\newcommand{\Rsun}{\hbox{$\hbox{R}_\odot\;$}}
\newcommand{\myemail}{s.repetto@astro.ru.nl}
\newcommand{\beq}{\begin{equation}}
\newcommand{\eeq}{\end{equation}}
\newcommand{\mr}{\mathrm}
\usepackage{amssymb,amsmath}
\usepackage{epsfig}
\usepackage{rotating}
\usepackage{times}
\usepackage{natbib,twoopt}
\usepackage[breaklinks=true, colorlinks=true]{hyperref} 
\hypersetup{colorlinks=True, linkcolor=BlueGreen}
\bibpunct{(}{)}{;}{a}{}{,}             
\makeatletter
  \newcommandtwoopt{\citeads}[3][][]{\href{http://adsabs.harvard.edu/abs/#3}%
    {\def\hyper@linkstart##1##2{}%
     \let\hyper@linkend\@empty\citealp[#1][#2]{#3}}}
  \newcommandtwoopt{\citepads}[3][][]{\href{http://adsabs.harvard.edu/abs/#3}%
    {\def\hyper@linkstart##1##2{}%
     \let\hyper@linkend\@empty\citep[#1][#2]{#3}}}
  \newcommandtwoopt{\citetads}[3][][]{\href{http://adsabs.harvard.edu/abs/#3}%
    {\def\hyper@linkstart##1##2{}%
     \let\hyper@linkend\@empty\citet[#1][#2]{#3}}}
  \newcommandtwoopt{\citeyearads}[3][][]%
    {\href{http://adsabs.harvard.edu/abs/#3}
    {\def\hyper@linkstart##1##2{}%
     \let\hyper@linkend\@empty\citeyear[#1][#2]{#3}}}
\makeatother

\title[The coupled effect of tides and stellar winds]
  {The coupled effect of tides and stellar winds on the
  evolution of compact binaries }
\author[S.~Repetto and G.~Nelemans]
  {Serena~Repetto$^{1}$\thanks{E-mail:
\myemail},~~Gijs~Nelemans$^{1,2}$\\
    $^1$Department of Astrophysics/IMAPP, Radboud University Nijmegen, P.O. Box 9010, 6500 GL Nijmegen, The Netherlands\\
    $^2$ Institute for Astronomy, KU Leuven, Celestijnenlaan 200D, 3001 Leuven, Belgium
  }

\date{Accepted XXX. Received XXX}
\pagerange{\pageref{firstpage}--\pageref{lastpage}} \pubyear{0000}
\pagerange{\pageref{firstpage}--\pageref{lastpage}} \pubyear{2014}
\def\LaTeX{L\kern-.36em\raise.3ex\hbox{a}\kern-.15em
    T\kern-.1667em\lower.7ex\hbox{E}\kern-.125emX}

\begin{document}

\label{firstpage}

\maketitle
\label{firstpage}

\begin{abstract}
We follow the evolution of compact binaries under the coupled effect of tides and stellar winds until the onset of Roche-lobe overflow. 
These binaries
contain a compact object
(either a black-hole, a neutron-star, or a planet)
and a stellar component.
We integrate the full set of tidal equations, which are based on Hut's model for tidal evolution,
and we couple them with the angular momentum loss in a stellar wind. Our aim is twofold. Firstly,
we wish to highlight
some interesting evolutionary outcomes of the coupling.
When tides are coupled with a non-massive stellar wind,
one interesting outcome
is that in certain types of binaries,
the stellar spin tends to reach a quasi-equilibrium state,
where the effect of tides and wind
are counteracting each other.
When tides are coupled with a massive wind, 
we parametrize the evolution in terms of
the decoupling radius, at which the wind decouples from the star. Even
for small decoupling radii this \emph{wind braking} can drive systems 
on the main sequence 
to
Roche-lobe overflow that otherwise would fail to
do so.
Our second aim is to inspect whether simple timescale considerations are a good description of the evolution of the systems. We find that simple timescale considerations, which rely
on neglecting the coupling between tides and stellar winds,
do not accurately represent the true evolution of compact binaries.
The outcome of the 
coupled evolution
of the rotational and orbital elements
can strongly differ from simple timescale considerations,
as already pointed out by
\citealt{2009MNRAS.395.2268B}
in the case of short-period
planetary systems.
\end{abstract}


\section{Introduction}
{
Just like the Earth and the Moon,
stars in binaries raise tidal bulges on each other.
Tidal interaction is an important ingredient in binary evolution,
and, on secular timescales,
it can drive the evolution of the system.
Tidal interaction between two bodies dissipates energy,
and induces exchange of angular momentum between the orbit and the spin of the components.
See \citealt{2008EAS....29...67Z}
for a recent comprehensive review
on tidal theory in binaries.

We study tidal interaction
in binaries formed
by a stellar component
and a companion which can be approximated as
a point mass,
i.e. neutron star (NS), black-hole (BH) or planet.
Unless the spin angular momentum of the stellar component
exceeds $1/3$ of the orbital angular momentum,
the binary orbital and rotational elements evolve in time,
and the binary is pushed towards the lowest-energy configuration,
on a timescale which depends on the strength of the tidal interaction.
This configuration
corresponds to a stable solution for the evolution
(\citealt{1973Ap&SS..23..459A}, \citealt{1879Obs.....3...79D}, \citealt{1980A&A....92..167H}).
The equilibrium is characterized
by circularity, tidal locking of the two components (synchronization), and 
alignment of the stellar spin with respect to the orbital spin.
However, an equilibrium solution
does not exist in case there is a sink of angular momentum in the system.
A possible sink is
any stellar wind which carries away angular momentum.
One type of angular-momentum loss we consider is
{\emph{magnetic braking}} {{(MB, see \citealt{1958ApJ...128..677P}, \citealt{1967ApJ...148..217W}, \citealt{1981A&A...100L...7V})}},
the loss of angular momentum in a magnetic stellar wind from a low-mass star.
The stellar wind,
being anchored to the magnetic field lines,
is forced to corotate out to large stellar radii,
causing large amount of angular momentum being lost from the star
and consequently from the orbit,
thanks to the tidal coupling.
Tides, counteracting the effect of magnetic braking,
induce a secular spiral-in of the two stars,
driving systems-
that would otherwise
remain detached- to Roche-lobe overflow (RLO).

Tidal evolution is to be considered in all of those systems which
acquire a certain 
degree
of eccentricity and/or asynchronism at some point in their evolution. For instance,
progenitors of black-hole and neutron-star X-ray binaries,
systems containing a star and a planet,
or binaries in globular clusters.
Tides are also important in one of the possible models for long gamma-ray bursts,
in which the black-hole progenitor
that undergoes
core-collapse
is required
to be a fast rotator.
A way of enhancing the rotation of the progenitor
is through synchronization
with the orbital motion
(see for example
\citealt{2012MNRAS.425..470C}).\\

The coupling between tides and magnetic braking in the progenitors of black-hole low-mass X-ray binaries (BH-LMXBs)
and neutron-star X-ray binaries (NS-LMXBs) has often been neglected in the past;
typically, one assumes that the binary rapidly circularizes
and synchronizes;
once synchronization is achieved,
every bit of angular momentum that is lost from the star,
is also lost from the orbit.
We aim at investigating whether this model is a good description
of the evolution of the systems.
In order to do so, we numerically integrate the full set of tidal equations,
coupled with magnetic braking,
until the Roche-lobe overflow configuration.
We then compare our results with the
estimates provided by the non-coupled method.
We do the same for binaries
containing a low-mass star
with a planetary companion.
We will extend our model to binaries
where the stellar companion is of high mass,
and we will see how the evolutionary equations need to be modified.

The effect of the coupling between tides and magnetic braking on the evolution
of short-period extrasolar planetary systems
was studied by
\citealt{2009MNRAS.395.2268B} 
(hereafter, BO2009). They found
that neglecting magnetic braking in this
type of systems
results in a very different outcome of the evolution. 
And in particular, they concluded that it 
is essential to consider the coupled evolution
of the rotational and orbital elements, 
a conclusion which was previously
pointed-out by 
\citealt{2008ApJ...678.1396J},
who also focused on the orbital
evolution of planetary systems.\\

This paper is divided into 6 sections:
in section \ref{sec:Model} we present our model
and we validate our code,
in section \ref{sec:binarieswestudy} we show the binaries we study
and 
we make some predictions of the evolution from simple timescale considerations,
in section \ref{sec:results} and \ref{sec:windbraking} we show our results,
and in section \ref{sec:discus} and \ref{sec:conclu} we 
discuss our results and we draw our conclusions.

\section{Model}
\label{sec:Model}
\subsection{Tides coupled with Magnetic Braking}
Magnetic braking is the loss of angular momentum 
in the stellar wind of a main-sequence magnetic star.
Low-mass stars
($0.3 \lesssim M_\star \lesssim 1.5~\Msun$)
have radiative cores and convective envelopes,
where magnetic fields are thought to 
be amplified by dynamo processes. Thanks to the corotation of the magnetic field lines out to large distances from the star,
any stellar wind takes away large amounts of angular momentum,
even if the mass-loss is negligible.
The braking is responsible
for the slow rotation
of most cool and old stars, like the Sun.
From Skumanich's empirical law (\citealt{1972ApJ...171..565S})
that describes
the dependence of the equatorial velocity
on age for main-sequence G stars,
the spin-down is obtained as:
\begin{equation}
\label{eq:MB}
\frac{d{\omega}_\star}{dt}=-\gamma_\text{MB} \text{R}_\star^2 \omega_\star^3
\end{equation}
where $\gamma_\text{MB}$ is measured as $\approx 5\times 10^{-29} \mr{s/cm}^2 $,
$ \text{R}_\star$ is the radius of the star,
and $\omega_\star$ its spin frequency.
For the radius of the star,
we use    
the mass-radius relation of a ZAMS-star as in 
\citealt{1996MNRAS.281..257T}.
(Note that 
the star can grow by 
up to a factor of 2 on the main sequence).
For a review on the different prescriptions
of magnetic braking see
\citealt{2011ApJS..194...28K}.

The star is doomed to lose its rotation
within its MS-lifetime,
unless it is tidally coupled to a companion:
the rotational angular momentum reservoir of the star is constantly being refueled 
by the tidal torque.

We follow the \citealt{1981A&A....99..126H} description of tidal interaction,
in which
tides have 
small deviations 
in magnitude and direction
from their 
equilibrium shape, 
and where dynamical effects are neglected.
{{In this model, the deviation of tides in magnitude and direction is
parametrized in terms of a constant and small time-lag $\tau$.}}
Orbital elements can therefore be assumed
to vary slowly within 
an orbital period, and 
their change 
can be computed through averages along
the orbit of the tidal potential.

We extend Hut's equations,
valid in the small-angle approximation,
to arbitrary inclinations of the stellar spin with the orbital spin,
using the same orbit-averaged approach.
{{We note that the tidal model by \citealt{1981A&A....99..126H}
was first extended to the case of arbitrary inclinations
by \citealt{1998ApJ...499..853E}.}}

Taking a binary with semi-major axis $a$,
spin frequency of the star $\omega_\star$,
eccentricity $e$,
and inclination of the rotational angular momentum
with respect to the orbital angular momentum $i$, 
we write the tidal equations for the evolution
of the rotational and orbital elements,
adding the magnetic braking term of equation \ref{eq:MB}:

\begin{eqnarray}
\label{eq:eq1}
\frac{da}{dt} && = -6\left (\frac{K}{T}\right )_i q (1+q)\left (\frac{R_\star}{a}\right )^8\frac{a}{(1-e^2)^{15/2}}\nonumber  \\
&& \left [f_1(e^2) -(1-e^2)^{3/2} f_2(e^2) \frac{\omega_\star \cos{i}}{\omega_\mr{orb}}\right ]\\
\frac{de}{dt} && = -27\left (\frac{K}{T}\right )_i q (1+q)\left (\frac{R_\star}{a}\right )^8\frac{e}{(1-e^2)^{13/2}}\nonumber  \\
\label{eq:eq2}
&& \left [f_3(e^2) -\frac{11}{8}(1-e^2)^{3/2} f_4(e^2) \frac{\omega_\star \cos{i}}{\omega_\mr{orb}}\right ]\\
\label{eq:eq3}
\frac{d\omega_\star}{dt} && = 3\Bigg (\frac{K}{T}\Bigg )_i \frac{q^2}{k^2} \Bigg (\frac{R_\star}{a}\Bigg )^6\frac{\omega_\mr{orb}}{(1-e^2)^{6}} \nonumber  \Bigg [ f_2(e^2)\cos{i}- \frac{1}{4}\frac{\omega_\star}{\omega_\mr{orb}} \nonumber  \\
&&  (1-e^2)^{3/2} (3+\cos{2 i}) f_5(e^2)\Bigg ] -\gamma_\text{MB} \text{R}_\star^2 \omega_\star^3 
\end{eqnarray}

\begin{eqnarray}
\label{eq:eq4}
\frac{di}{dt}&&= -3\left (\frac{K}{T}\right )_i \frac{q^2}{k^2} \left (\frac{R_\star}{a}\right )^6\frac{\sin{i}}{(1-e^2)^{6}} \frac{\omega_\mr{orb}}{\omega_\star}\Bigg [ f_2(e^2) -\nonumber  
\frac{f_5(e^2)}{2} \nonumber \\
&& \times \Bigg (\frac{\omega_\star \cos{i} ~(1-e^2)^{3/2}}{\omega_\mr{orb}} +  
\frac{R_\star^2  a~\omega_\star^2 k^2 (1-e^2)}{M_\mr{comp} G} \Bigg ) \Bigg ]
\end{eqnarray}
where $M_\mr{comp}$
is the mass of the compact companion, 
$M_\star$ the mass of the star,
$q$ the mass-ratio $M_\mr{comp}/M_\star$,
and the functions $f_i(e^2)$ are polynomials in 
$e^2$ as in \citealt{1981A&A....99..126H}.
For the radius of gyration $k$,
we use the fitting formula given in \citealt{2013ApJ...764..166D}, which is based on the detailed stellar evolution models of \citealt{1998MNRAS.298..525P},
and which gives the mass-dependence of $k$.\\

The calibration factors $\left \{K/T\right \}_{i=c,r}$
measure the strength of
the dissipation of the tidal flow.
{{$K$ is the apsidal motion constant,
which takes into account the central condensation of the star
(\citealt{1976ApJ...205..556L}). $T$ is a typical timescale
on which significant changes in the orbit take place
through tidal evolution;
in units of the orbital period, $T$ is the inverse of the
tidal time-lag $\tau$. }}
{{In his derivation,
Hut considers a constant time-lag $\tau$. However,
it has been later argued whether this was the appropriate choice.
The time-lag should in fact be compared
with typical relaxation time
of the process responsible for the dissipation.}}
The source of dissipation
depends on the type of star which
undergoes tidal deformations (see \citealt{1977A&A....57..383Z}).

Low mass stars have convective envelopes and it is believed that turbulent convection
is responsible for the dissipation.
{{When viscosity results from turbulence,
the natural relaxation time
is the eddy-turnover timescale $\tau_\mr{conv}$. This may be longer than the orbital period,
in which case the efficiency of the dissipation should be reduced.
The efficiency should then be dependent on the tidal forcing frequency
(see \citealt{1977Icar...30..301G}, \citealt{1989A&A...220..112Z}, \citealt{2004MNRAS.353.1161I}).}}

{{Two scalings have been proposed
for the viscosity due to the turbulent convection. 
\citealt{1966AnAp...29..489Z} assumes a linear scaling with
the tidal forcing frequency.
He expresses $f_\mr{conv}$,
the fraction of the convective cells which contribute to the damping, as:
\beq
\label{eq:fconvzahn}
f_{\mr{conv}} = \min \left[ 1 , \left( \frac{P_{\rm tid}}{2 {\tau}_{\mr{ conv}}} 
\right) \right]
\eeq
}}
where $P_{\mr{ tid}}$ is the tidal forcing period, given by:
\beq
\frac{1}{P_{\mr{tid}}} = \left | \frac{1}{P_{\mr{ orb}}} - \frac{1}{P_{\mr{\star}}}
\right | 
\eeq
For high tidal-forcing frequency (i.e. when $P_\mr{tid} \ll \tau_\mr{conv}$)
the efficiency of momentum transfer
by the largest convective cells
is reduced. 
{{Instead, \citealt{1977Icar...30..301G}
suggested a quadratic dependence:}}
\beq
\label{eq:fconv}
f_{\mr{conv}} = \min \left[ 1 , \left( \frac{P_{\rm tid}}{2 {\tau}_{\mr{ conv}}} 
\right)^2 \right]
\eeq
{{There is a long-standing uncertainty regarding
which scaling is correct,
with some numerical simulations
of turbulent viscosity in a convection zone
favoring a linear scaling (\citealt{2007ApJ...655.1166P}),
while others favoring a quadratic scaling
(\citealt{2012MNRAS.422.1975O}).

{{The question of tidal dissipation in a convective shell
is still ``Achille's heel of tidal theory'', as \citealt{2008EAS....29...67Z} wrote,
and it goes beyond the scope of this Paper.
We choose $f_\mr{conv}$ as in eq. \ref{eq:fconv}.
In this we follow
the approach by 
\citealt{2002MNRAS.329..897H}, \citealt{2008ApJS..174..223B} and \citealt{2014ApJ...786..102V},
who express the calibration factor
for a convective envelope
as: }}
}}
\begin{equation}
\label{eq:calibconv}
\left (\frac{K}{T} \right )_c=\frac{2}{21} \frac{f_\mr{conv}}{\tau_\mr{conv}}\frac{M_\mr{env}}{M_\star}~[\mr{yr}^{-1}]
\end{equation}
where 
$M_\mr{env}$ is the mass in the convective envelope.

{{Zahn's prescription
seem to be in better in agreement with observations
of tidal circularization times for binaries containing a giant star
(\citealt{1995A&A...296..709V}).
}}
{{A more recent work by \citealt{2008ApJS..174..223B},
showed the need for
multiplying $f_\mr{conv}$ in eq. \ref{eq:fconv}
by $50$, to match the observed circularization period
of close binary stars. However, due to general uncertainties
on tidal calibration for a convective envelope,
we find our model a reasonable place to start.

{{Another uncertainty is whether and how
the time-lag should depend on the misalignment between the stellar spin and the orbital angular 
momentum. There is not yet agreement on this (Adrian Barker, priv. communication),
and we prefer not to add additional parameters to our model.}}

}}

High-mass stars have radiative envelopes,
and tidal motions
are assumed to be dissipated
via radiative damping of the stellar $g$-modes. 
In this case
the calibration factor is (\citealt{2002MNRAS.329..897H}, \citealt{2008ApJS..174..223B}):
\beq
\label{eq:KTrad}
\left (\frac{K}{T} \right )_r = 1.9782 \times 10^4 \frac{M_\star R_\star^2}{a^5} 
\left( 1 + q \right)^{5/6} E_2 \: [{\rm yr}^{-1}] 
\eeq
where $E_2$
is a second-order parameter which measures the coupling between the tidal potential and the gravity mode
(\citealt{1975A&A....41..329Z});
we fit it to the values given by \citealt{1975A&A....41..329Z}:
\beq 
E_2 = 1.58313 \times 10^{-9} M_\star^{2.84} 
\eeq 
(see also \citealt{2002MNRAS.329..897H}).

\subsection{Non-coupled methods}
One way of following the evolution
of a binary system
in which 
tidal friction and magnetic braking
are both operating,
is to assume synchronization
and 
to circularize the binary
instantaneously or
on a certain circularization timescale,
neglecting the spin of the star. The new semi-major axis
is then $a_\mr{circ}=a(1-e^2)$.
This choice appears reasonable when dealing
with systems in which the angular momentum
stored in the tidally-deformed component is small
(i.e. the moment of inertia of the star is small).
{{Tidal interaction conserves the total angular momentum
$J_\mr{tot}=J_\mr{orb}+J_\star$.
Neglecting $J_\star$,
conservation of $J_\mr{orb}$
gives $a_\mr{circ}=a(1-e^2)$.}}
Afterwards,
the tidal torque
counteracts the effect of the magnetic spin-down,
bringing angular momentum from the orbit
back into the star. It is then reasonable to assume that
every bit of angular momentum which is lost from the star,
is also lost from the orbit (i.e. $\dot{J}_\mr{orb}=\dot{J}_\star$,
see \citealt{1981A&A...100L...7V}).
The rate of
angular momentum loss in the stellar wind is:
\begin{equation}
\frac{dJ_\star}{dt}=k^2 M_2 R^2_\star \frac{d{\omega}_\star}{dt}
\end{equation}
where
we have neglected the change in radius
and mass of the star, 
for the mass-loss being negligible
(typically, $|\dot{M}_\star| \sim10^{-14}~\Msun\mr{/yr}$ for a Sun-like star).
Using $\dot{a}/{a} = {2\dot{J}_{\mr{orb}}}/{J_\mr{orb}}={2\dot{J}_\star}/{J_\mr{orb}}$ and
equation \ref{eq:MB},
and assuming $\omega_\star=\omega_\mr{orb}$,
the semi-major axis decay rate is:
\begin{equation}
\label{eq:MBonTheBinary}
\frac{da}{dt}=- \frac{2 \gamma_\mr{MB} k^2 R_\star^4 G M^2}{M_\mr{comp}} \frac{1}{a^4}
\end{equation}
where $M$ is the total mass of the binary.
We will call this approach \emph{non-coupled} method.

Our approach instead is to numerically-integrate the full set of tidal equations
coupled with the magnetic spin-down
(eqs. \ref{eq:eq1}-\ref{eq:eq2}-\ref{eq:eq3}-\ref{eq:eq4}).
In section \ref{sec:results} we present the results of this
integration for some illustrative binaries,
showing how they differ from
the results in the non-coupled method.
A particular stress will be given to circularization timescales.
Some previous works used these timescales 
as a simple way of depicting the evolution of compact binaries.
It turns out that the true evolution
of the eccentricity can
highly differ from the exponential decay 
implied by the choice of a circularization timescale
as $e/\dot{e}$.

\subsection{Validation of the evolution code}
For small values of the inclination,
our tidal equations for an eccentric and non-coplanar binary
recover the equations in \citealt{1981A&A....99..126H}.
In contrast to the BO2009
equations,
we write down explicitly the
tidal equations for $(a, e, \omega_\star, i)$
in the case of a non-circular and non-coplanar orbit.
BO2009 choose instead of constant time-lag $\tau$ like in Hut's model,
a constant $Q^\prime$,
where $Q^\prime$ is 
a re-parametrization
of the tidal calibration factor $Q$.
The factor $Q$,
also called \emph{quality factor},
is a dimensionless quantity used
mainly in planetary studies
for characterizing
the efficiency of tidal dissipation.
{{BO2009 take $Q^\prime=10^6$
(see also \citealt{2007ApJ...661.1180O}).}}
We recover the equations 4, 5, 6 in BO2009
for a circular and non-coplanar orbit,
when replacing
$K/T$
in our equations
with $\frac{3}{2Q^\prime} \frac{1}{\omega_\mr{orb}} \frac{G M_\star}{R_\star^3}$.

In order to validate our code,
we reproduce some of the results presented by
BO2009.
Figure \ref{fig:Fig1Barker}
is a phase-portrait plot,
showing the evolution
in the plane ($\tilde{n}$, $\tilde{\Omega}$)
of coplanar and circular
systems with varying initial conditions.
{{The variables 
$\tilde{n}$ and $\tilde\Omega$
are 
re-parametrization
of the orbital frequency
and the stellar spin frequency respectively,
which have been normalized to 
the orbital frequency at the stellar surface,
together with a constant factor (see BO2009 for details).}}
In integrating our evolutionary equations,
we have replaced our tidal calibration factor $K/T$
with $\frac{3}{2Q^\prime} \frac{1}{\omega_\mr{orb}} \frac{G M_\star}{R_\star^3}$,
and we have used BO2009 
calibration for magnetic braking
($\gamma_\mr{MB}$
in our model is
$0.4$ times smaller than $\gamma_\mr{MB}$ in BO2009).
Figure \ref{fig:Fig1Barker} coincides with the top plot
in fig.1 of BO2009.
Once magnetic braking spins down the star below corotation
($\tilde{\Omega}<\tilde{n}$)
the orbit undergoes tidally-induced decay.

We also integrate our 
evolutionary equations
for a set of planetary systems
with circular and non-coplanar orbit
($i=90^\circ$),
and we show the corresponding
solutions
in the phase-portrait
plot of figure \ref{fig:Fig3Barker}.
The overall evolution
is similar to the one in figure \ref{fig:Fig1Barker},
the only difference being that
system with $\tilde\Omega \cos{i} < \tilde{n}$
undergo orbital decay 
while still being outside corotation.
This plot is identical to
the top plot in fig.3 in BO2009.

\begin{figure*}
\begin{minipage}{\columnwidth}
\centering
\includegraphics[width=\textwidth]{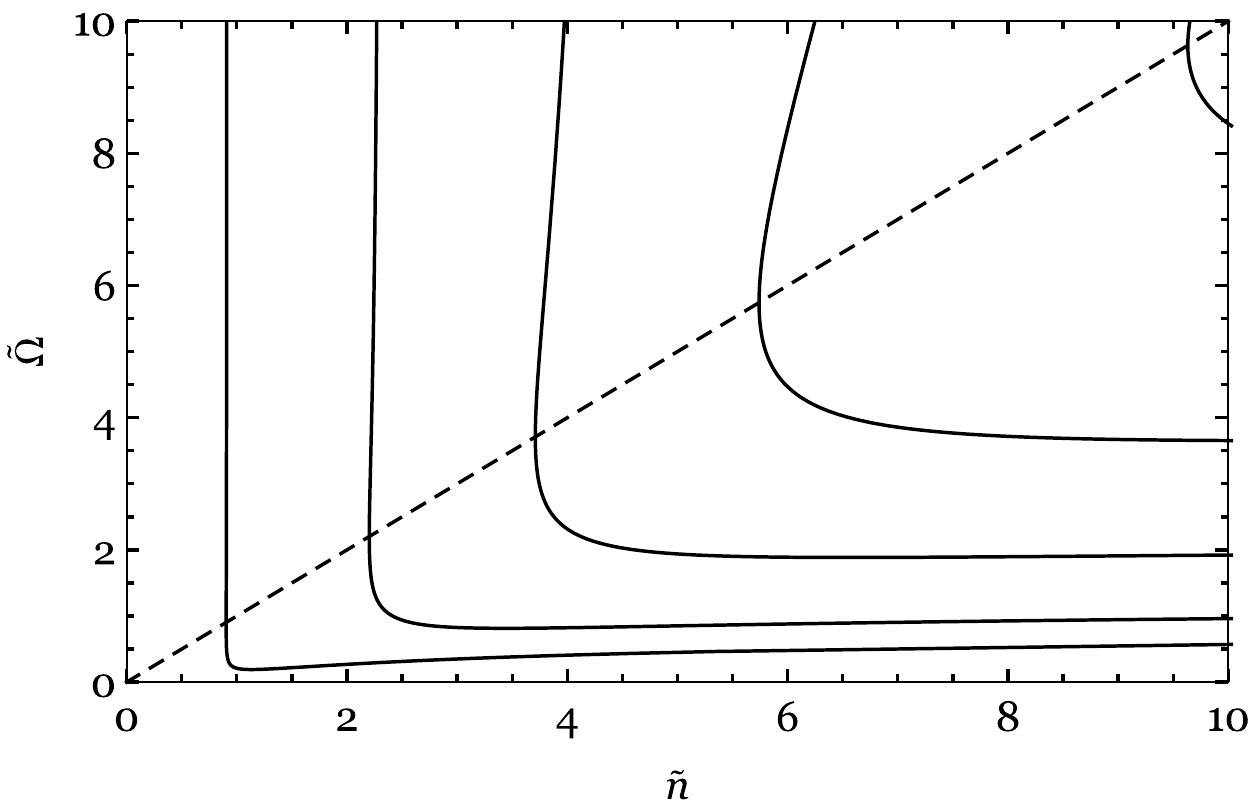}
\caption{Evolution
of a coplanar and circular planetary system with varying initial conditions
in the plane ($\tilde{n}$, $\tilde{\Omega}$) (solid lines).
The diagonal line corresponds 
to corotation $\tilde{\Omega}=\tilde{n}$.}
\label{fig:Fig1Barker}
\end{minipage}
\hspace{0.3cm}
\begin{minipage}{\columnwidth}
\centering
\includegraphics[width=\textwidth]{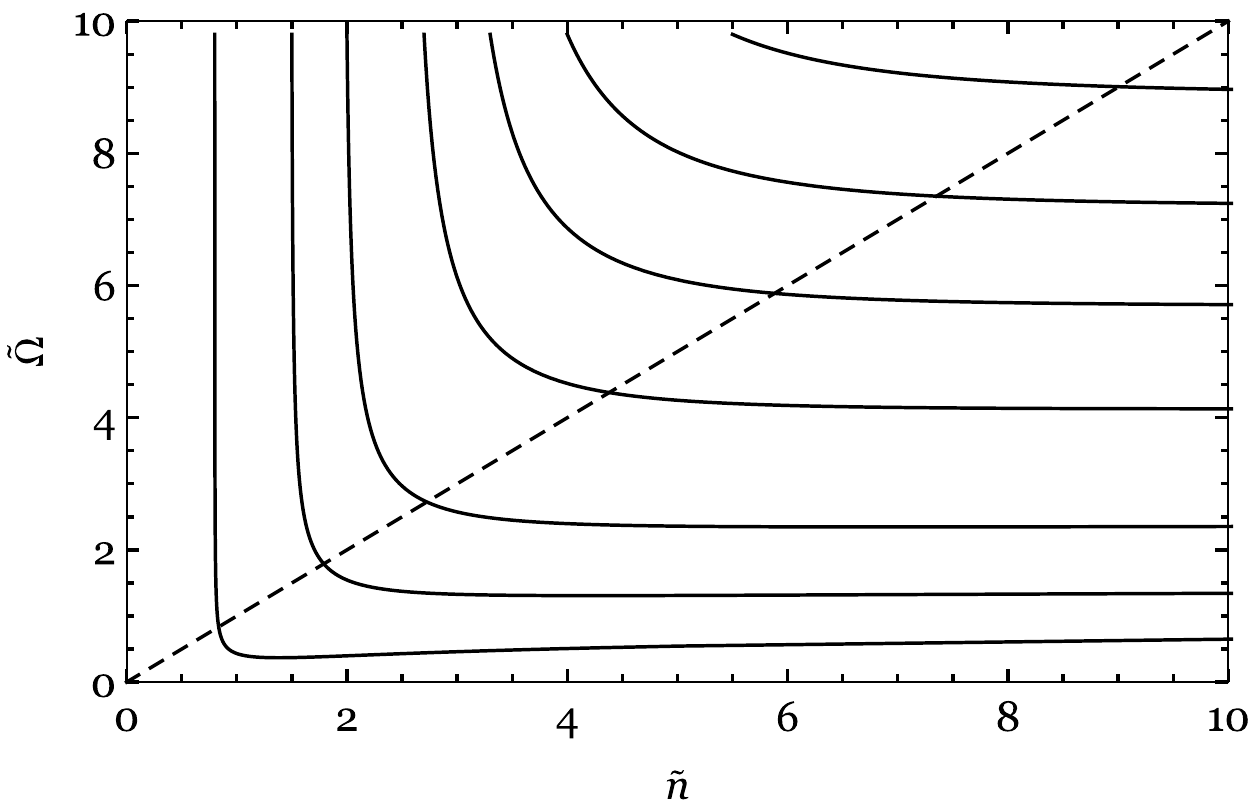}
\caption{Evolution
of a circular and {{non-coplanar ($i=90^\circ$)}} planetary system with varying initial conditions
in the plane ($\tilde{n}$, $\tilde{\Omega}$) (solid lines).
The diagonal line corresponds 
to corotation $\tilde{\Omega}=\tilde{n}$.}
\label{fig:Fig3Barker}
\end{minipage}
\hspace{0.3cm}
\end{figure*}

\section{Binaries in our study}
\label{sec:binarieswestudy}
We present the binaries we consider in our study
in figure \ref{fig:Overall}. 
The most extreme ones are binaries
in which the magnetic wind from a low-mass star
is coupled with the tidal friction
between the star and its compact companion.
The companion 
is either a black-hole,
or a Hot Jupiter,
{{which is a planet with mass similar to Jupiter 
at a distance less than $0.1$ AU
from its host star.}}

We will also extend our model
to a case in which the star is
of high-mass instead (see section \ref{sec:windbraking}). The companion is
either a neutron star,
a black-hole,
or, in a more exotic scenario,
a white dwarf (WD).

Together these systems
span a wide range of mass ratios,
ranging from
$q=0.001-10$.\\

We will refer to a binary
containing a black hole or a neutron star
as a black-hole X-ray binary of high or low mass (BH-HMXB/BH-LMXB),
or a neutron-star high/low-mass X-ray binary (NS-HMXB/NS-LMXB),
where low and high refer
to the mass of the companion.
We point out though 
that these terms commonly refer to the 
mass-transfer phase,
which is one of the evolutionary stages 
experienced by these binaries.
What we deal with instead
are systems on their way
of becoming X-ray binaries,
that is, we study
the secular phase before mass-transfer sets in.

\begin{figure}
\centering
\includegraphics[width=\columnwidth]{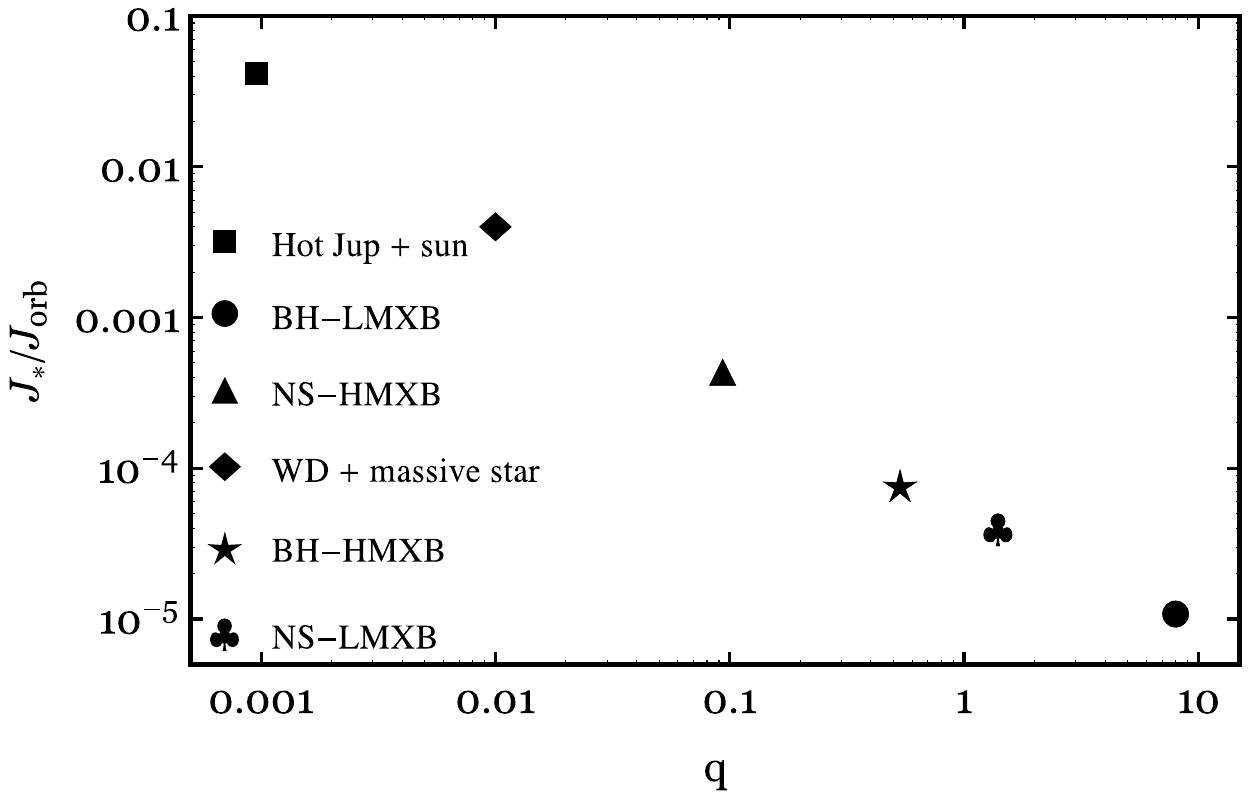}
\caption{Different types of compact binaries as a
 function of the ratio $J_\star/J_\mr{orb}$ between the rotational and the orbital angular momentum and as a function of the binary mass ratio q.}
\label{fig:Overall}
\end{figure}

\subsection{Timescale considerations}
\label{sec:timescales}
Before going into the details of 
the numerical integration of the coupled equations
\ref{eq:eq1}-\ref{eq:eq2} -\ref{eq:eq3} -\ref{eq:eq4},
we wish to make some predictions
on the evolution of the systems
based on simple timescale considerations \footnote{As a note, throughout the paper we
will indicate the timescale of a certain event
with $\tau$,
whereas $t$ will indicate
the time of occurrence of the same event 
in our integration.
}.

The binaries in our study are presented 
in figure \ref{fig:Overall} in terms of their
mass-ratio $q$ and 
the ratio $\eta=J_\star/J_\mr{orb}$. $J_\mr{orb}$ is calculated for a circular 
orbit with $a=2~a_\mr{RLO}$.
$a_\mr{RLO}$ is the
orbital separation at which the system undergoes RLO,
i.e. $a_\mr{RLO}=R_\star/f(1/q)$,
where $f(q)$ is a function of the mass-ratio
as in
\citealt{1983ApJ...268..368E}.
$J_\star$ is calculated 
for a slow-spinning star with
$\omega_\star=10^{-3}~\omega_\mr{break}$,
where $\omega_{\mr{break}}$ is the break-up frequency $\sqrt{{G M_\star}/{R_\star^3}}$.
The ratio $\eta$ is then larger 
for faster-spinning star.
It is
an indicator for how much 
angular momentum is stored in the star
compared to how much is stored in the orbit.
A larger $\eta$,
like in the planetary system case,
implies 
that tides will more easily bring changes to the 
orbit than to the spin.
If the orbit is sufficiently close,
the planetary system
suffers a significant tidal decay,
while the tidal change of the 
stellar spin is much less significant. 
The opposite happens
in a BH-LMXB,
where tides easily affect the stellar spin.

The larger the mass-ratio,
the smaller the timescale
on which tides
cause significant changes to the orbital and rotational elements,
as can be seen from the dependence on the mass-ratio 
of the tidal equations \ref{eq:eq1}-\ref{eq:eq2}-\ref{eq:eq3}-\ref{eq:eq4}.
Hence tides are much more efficient
in a BH-LMXB than
in a planetary system.\\

We can obtain a rule of thumb
for determining which systems will first circularize rather than synchronize.
The circularization timescale can be estimated as ${e}/{\dot{e}}$
where $\dot{e}$ is as in equation \ref{eq:eq2},
and assuming $e\approx 0$ and $\omega_\star=\omega_\mr{orb}$
(\citealt{2002MNRAS.329..897H}):
\beq\label{eq:tau_circ}
\frac{1}{{\tau}_{\rm circ}} =  
\frac{21}{2} \left( \frac{K}{T} \right)_{\rm c} q \left( 1 + q \right) 
\left( \frac{R_\star}{a} \right)^8 
\eeq
The synchronization timescale for a circular orbit is calculated
as (\citealt{2002MNRAS.329..897H}):
\begin{equation}
\label{eq:tau_sync}
\frac{1}{\tau_\mr{sync}}=\left | \frac{\dot{\omega}_{\star | \mr{tid}}}{\omega_\star-\omega_\mr{orb}}\right |
= 3 \left (\frac{K}{T}\right ) \frac{q^2}{k^2} \left (\frac{R_\star}{a}\right )^6
\end{equation}
Then $\tau_\mr{circ} < \tau_\mr{sync}$ when:
\begin{equation}
\frac{k R_\star}{a} > \sqrt{\frac{6}{21} \frac{q}{1+q}}
\end{equation}
where the star can be modeled
as a rigid sphere of radius $k R_\star$.
Taking $a=2~a_\mr{RLO}$,
the criteria is satisfied
for two of the types of binaries we showed in figure \ref{fig:Overall}:
the planetary system
and the binary consisting of a white-dwarf
and a massive star.
In all the other cases,
we expect the
first phase of the evolution
of the binary to be driven by changes in the stellar spin
rather than the orbit.\\

In our model the spin is being affected both by tides and
by magnetic braking;
we compare their typical timescales.

The MB spin-down timescale is calculated as:
\begin{equation}
\label{eq:MBspindown}
\frac{1}{\tau_\mr{MB}}=\frac{\dot{\omega}_{\star |{\mr{MB}}}}{\omega_\star}
\end{equation}
where $\dot{\omega}_{\star |{\mr{MB}}}$ is as in equation \ref{eq:MB}.
In figure \ref{fig:figure1},
we show the synchronization timescale
and the circularization timescale 
for a binary composed of a low-mass star
$\mr{(}M_\star=1~\Msun\mr{)}$,
as a function of the compact companion mass (solid black and solid gray line).
Timescales are calculated
using equations \ref{eq:tau_circ} and \ref{eq:tau_sync}.
The orbital separation 
is taken to be $a=2~a_\mr{RLO}$
and the star is spinning
with $\omega=0.9~\omega_\mr{break}$.
We also show
the MS-lifetime,
the MB spin-down timescale 
for a star spinning at $0.9~\omega_\mr{break}$,
the MB spin-down timescale
for a star spinning at $\omega_\mr{eq}$,
and for a star spinning at $10^{-5}~\omega_\mr{break}$.

{{The quasi-equilibrium frequency $\omega_\mr{eq}$}}
is the frequency at which MB and tidal torque
are balancing each other (i.e. $\left |\dot{\omega}_\star{_\mr{,MB}}\right |=\left | \dot{\omega}_\star{_\mr{,tid}}\right |$, see section
\ref{sec:results} for further details).
It has been calculated
for an orbit with $a=2~a_\mr{RLO}$
and $e=0.1$.

For low mass-ratios,
tides are weak.
Consequently, magnetic braking spins-down the star below corotation ($\omega_\star<\omega_\mr{orb}$), and tides won't manage to synchronize 
the spin frequency to the orbital frequency 
within a MS-lifetime.
This is what happens in the planetary-system case.
Instead, in a BH-LMXB
tides and magnetic braking 
tend to reach a {{quasi-equilibrium state}}
in which they balance each other.

Concerning the circularization timescale,
it is typically shorter than the synchronization timescale
for $q<0.05$,
like in the planetary system case.

When tides and magnetic braking are coupled, 
we can use this figure
to predict the evolution of the binary.
If a star is initially spinning at $0.9~\omega_\mr{break}$,
$\tau_\mr{MB}$ is shorter
than $\tau_\mr{sync}$.
This means that in the 
first phase of the evolution MB spins down the star,
until the timescale on which tides act on the spin
becomes comparable to the MB timescale,
and tides start playing a role.\\

\begin{figure}
\centering
\includegraphics[width=\columnwidth]{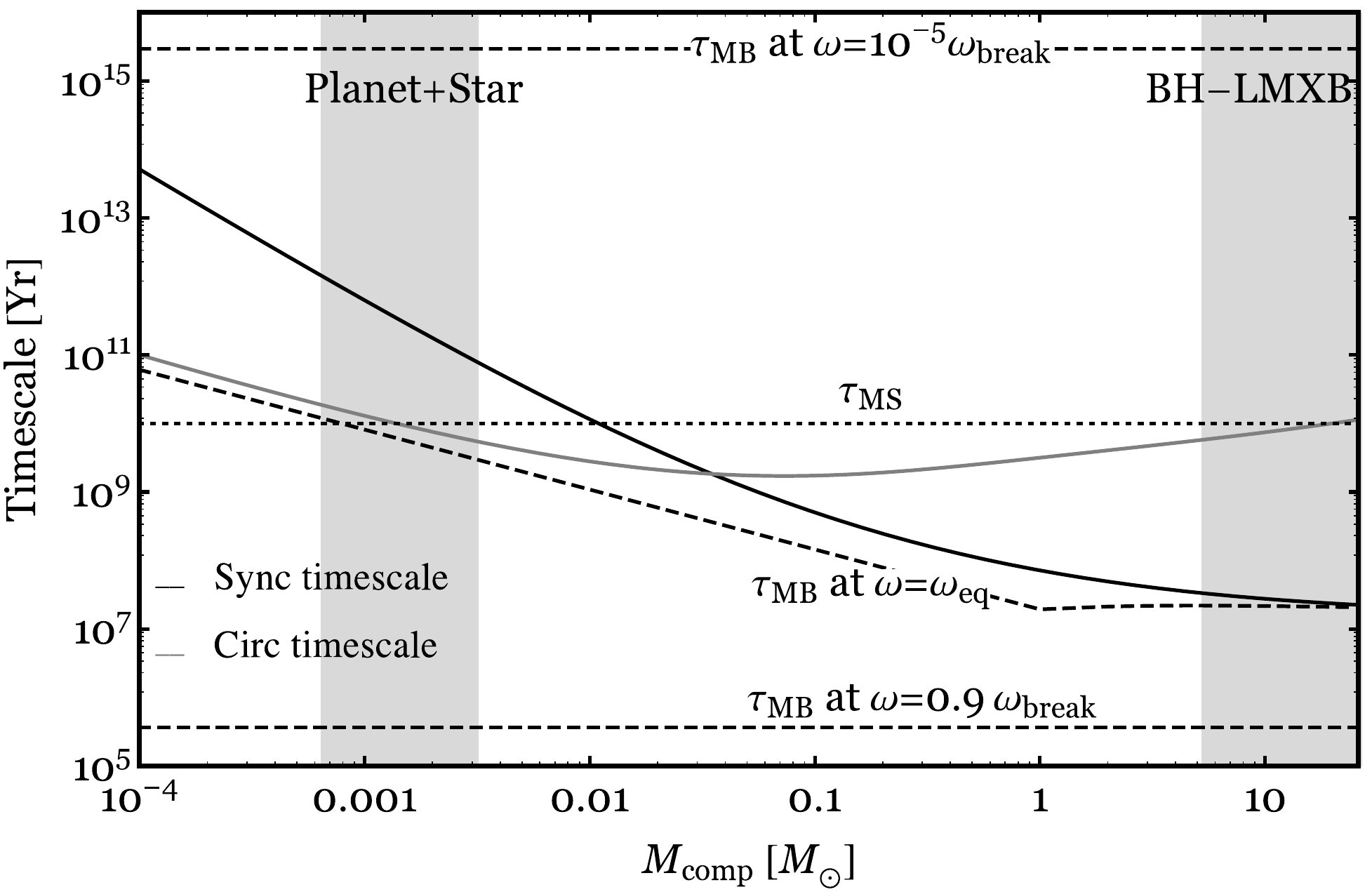}
\caption{The solid black line
and the solid gray line
show respectively the synchronization and
circularization timescale for a binary with $a=2~a_\mr{RLO}$ composed of a sun-like star
and a companion of mass $M_\mr{comp}$.
The dotted line shows the MS-lifetime, whereas the dashed lines show the MB spin-down timescale for $\omega_\star=0.9~\omega_\mr{break}$,
the MB spin-down timescale for $\omega_\star=\omega_\mr{eq}$ 
(see text for details on $\omega_\mr{eq}$),
and for $\omega_\star=10^{-5}~\omega_\mr{break}$.}
\label{fig:figure1}
\end{figure}

\section{Results}
\label{sec:results}
\subsection{Black-hole low-mass X-ray binary}
The orbit of a black-hole low-mass X-ray binary is
typically eccentric right after the formation of the black-hole.
The eccentricity is the result of the mass ejection and/or possibly the natal kick in the supernova
which gives birth to the black hole.
In addition, any component of the
natal kick
perpendicular to the orbital plane
will result in a misalignment between
the orbital spin
and the stellar spin.
The most effective way of shrinking a BH-LMXB down to the RLO configuration 
is a coupling between the magnetic wind from the low-mass star
and tidal friction,
whereas gravitational wave emission can be neglected
for the typical initial orbital separations ($a \approx 10~R_\odot$).

Previous binary population synthesis (BPS) works
on the evolution 
of BH-LMXBs
have neglected the coupling between
the stellar spin and the orbital spin,
following the evolution
of the binaries
according to what we call the non-coupled
method
(see for example
\citealt{1999ApJ...521..723K}
and \citealt{2006A&A...454..559Y}).
This has been done
for NS-LMXBs too,
see for example
\citealt{1988A&A...191...57P}
and \citealt{2009ApJ...691.1611M}.
This choice was motivated
by the fact
that the rotational angular momentum is small
compared to the orbital angular momentum.
Also, the tidal evolution of the misalignment
between the spin of the star
and the orbital spin has usually been neglected.\\

\subsubsection{An illustrative example of the evolution in the coplanar case}
We integrate
the tidal equations \ref{eq:eq1}, \ref{eq:eq2}, \ref{eq:eq3}, \ref{eq:eq4}  for an illustrative system formed
by a black hole of mass $M_\text{BH}=8~M_\odot$ and a star of mass $M_\star=1~M_\odot$,
until the RLO configuration is reached
($a_\mr{RLO}\approx 4~\Rsun$).
The 
rotational angular momentum is small
compared to the orbital one (see
figure \ref{fig:Overall}),
thus we expect the 
system to rapidly reach the
synchronous state.

\begin{figure*}
\begin{center}
\includegraphics[width=\textwidth]{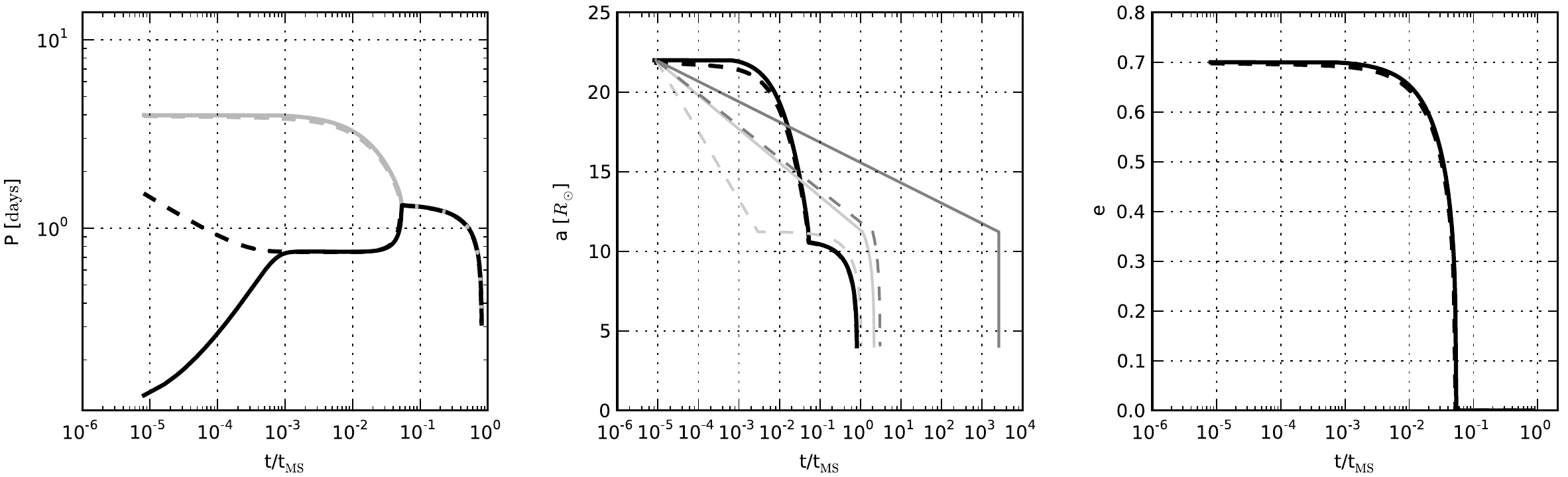}
\caption{Evolution of a coplanar and eccentric BH-LMXB under the effect of tides and magnetic braking.
The binary is composed of a black-hole of mass $8~M_\odot$
and a one solar-mass star. Initial orbital parameters are
$a=22~R_\odot$, $e=0.7$, $i=0$.
We consider both a high-spinning star ($\omega_\star=
0.9~\omega_\mr{break}$) and
a low-spinning star ($\omega_\star=10^{-5}~\omega_\mr{break}$).
In all the three panels,
the solid lines correspond to the high-spin,
the dashed lines to low-spin.
The left panel shows the evolution
of the stellar rotation period in black
and of the orbital period in grey.
The middle panel shows the evolution of the orbital separation,
whereas the right one represents the evolution of the eccentricity.
In the center panel, 
the two thick black lines represent the evolution
of the orbital separation in our integration.
The thin dark-gray lines
represent the evolution of the orbital separation
in the non-coupled method
when taking $\tau_\mr{circ}=\left .\left ({e}/{\dot{e}}\right )\right\vert _\mr{e\approx0, \omega_\star=\omega_\mr{orb}}$;
thin light-gray lines when taking $\tau_\mr{circ}=e/\dot{e}$.
}
\label{fig:bhstar}
\end{center}
\end{figure*}

In figure \ref{fig:bhstar} we show the evolution 
for initial orbital parameters
$a=22~R_\odot$, $e=0.7$ and $i=0$. Concerning the spin of the star,
we consider both a star spinning
at $\omega_\star= 0.9~\omega_{\mr{break}}$
(see solid thick lines),
and at $\omega_\star=10^{-5}~\omega_\mr{break}$ (dashed thick lines).

Either via magnetic spin-down 
(high-spin case) or via tidal spin-up (low-spin case),
the spin frequency converges to 
$\omega_\text{eq}$,
where $\omega_\text{eq}$ is the spin-frequency such
that 
$\left |\dot{\omega}_\star{_\mr{,MB}}\right |= \left |\dot{\omega}_\star{_\mr{,tid}}\right |$
(plateau in the left panel in figure \ref{fig:bhstar}).
{{The approach of this \emph{quasi-equilibrium} state \footnote{We call it \emph{quasi-equilibrium}
state
due to the fact that the orbital properties 
are still evolving.
}}}
is allowed by the fact that both 
${\dot{\omega}_{\star\mr{,MB}}}$
and ${\dot{\omega}_{\star\mr{,tid}}}$
depend on the stellar spin (see
equations \ref{eq:MB} and \ref{eq:eq3}).

Both solutions reach the quasi-equilibrium state $\omega_\text{eq}$
on a similar timescale,
which is very short ($\approx 10^{-3}~t_\mr{MS}$),
therefore no significant changes in the orbit take place during this phase.
Afterwards, the orbit circularizes and the spin
synchronizes to the orbital frequency.
We note that the 
time for the orbit to become circular
is non-negligible 
($t_\mr{circ}\sim 6\times10^{-2}~t_\mr{MS}$).
Once synchronization and circularization are achieved,
every bit of angular momentum which is lost from the star in the wind,
is also lost from the orbit.
The two components effectively approach each other until RLO,
and equation
\ref{eq:MBonTheBinary}
describes well the evolution of the system.
During this phase of the evolution,
tides and magnetic braking 
are perfectively
counteracting each-other,
i.e.
$\dot{\omega}_\star{|_\mr{MB}}= \dot{\omega}_\star{|_\mr{tid}}$.

\subsubsection{Comparison with non-coupled methods}
We also compare
the results of our integration
with the estimates of the non-coupled method,
both for the low-spinning case
and the high-spinning case. 
The results are presented in
the center panel of figure \ref{fig:bhstar},
where we show the evolution of the semi-major axis
in the non-coupled method. 
We take two types of circularization
timescale;
one as in equation \ref{eq:tau_circ} (dark gray lines),
and one given by $\tau_\mr{circ}=e/\dot{e}$ (light gray lines).
Solid lines are for the high spinning case
and dashed lines for the low spinning case.
The circularized orbital
separation is overestimated in both scenarios,
since changes in the orbit 
once the binary reaches
the quasi-equilibrium state,
are neglected in non-coupled methods.
As a consequence, 
the age of the binary at the RLO configuration 
is overestimated.
When taking $\tau_\mr{circ}=e/\dot{e}$,
 it is overestimated
by a factor of $\approx 2.6$ in the high-spin case,
and by $\approx1.3$ in the low-spin case.
The time it takes to reach circularization
in our integration is $t_\mr{circ}\approx 0.06~t_\mr{MS}$.
The ratio $t_\mr{circ}/\tau_\mr{circ}$ when taking
$\tau_\mr{circ}=e/\dot{e}$, is
$0.05$ in the high-spin case,
and $20$ in the low-spinning case.
With $\tau_\mr{circ}$
as in equation \ref{eq:tau_circ},
the ratio is $2\times 10^{-5}$
and $0.03$ respectively.

A better estimate of the time it takes to reach circularization
would arise taking  
a star spinning at half the break-frequency
in the calculation of $e/\dot{e}$
(see for details 
figure \ref{fig:RatioTimescalesBH+star}).

This illustrative example
highlights the importance, 
at least in the first phase of the evolution
(before circularization is achieved),
of considering the coupled evolution
of the rotational and orbital element.

\subsubsection{An illustrative example of the evolution in the misaligned case}
Misalignment has usually been neglected, and
we wonder whether it brings significant changes
to the orbital evolution.
Due to the small angular momentum stored in the star
($J_\star/J_\mr{orb} \ll 1$),
we expect the tidal torque to easily affect $J_\star$.
That is, we expect the stellar spin to rapidly
align and synchronize with the orbital spin.

In figure \ref{fig:BH-LMXBmisaligned}
we show a representative illustration of the evolution of a BH-LMXB
with initial orbital parameters $a=9~R_\odot$, $e=0.5$, 
with the star in a retrograde orbit around the black hole ($i=120^\circ$) and
spinning at $0.9~\omega_\mr{break}$.
Once magnetic braking
has spun-down the star sufficiently,
the spin aligns with the orbital angular momentum.
This can be seen in the right panel in the figure,
which shows how steeply the inclination
decreases once the spin of the star has become negligible.
An interesting difference with respect to the evolution
in the coplanar case,
is the initial decay of the orbital separation
of a retrograde orbit.
\begin{figure*}
\includegraphics[width=\textwidth]{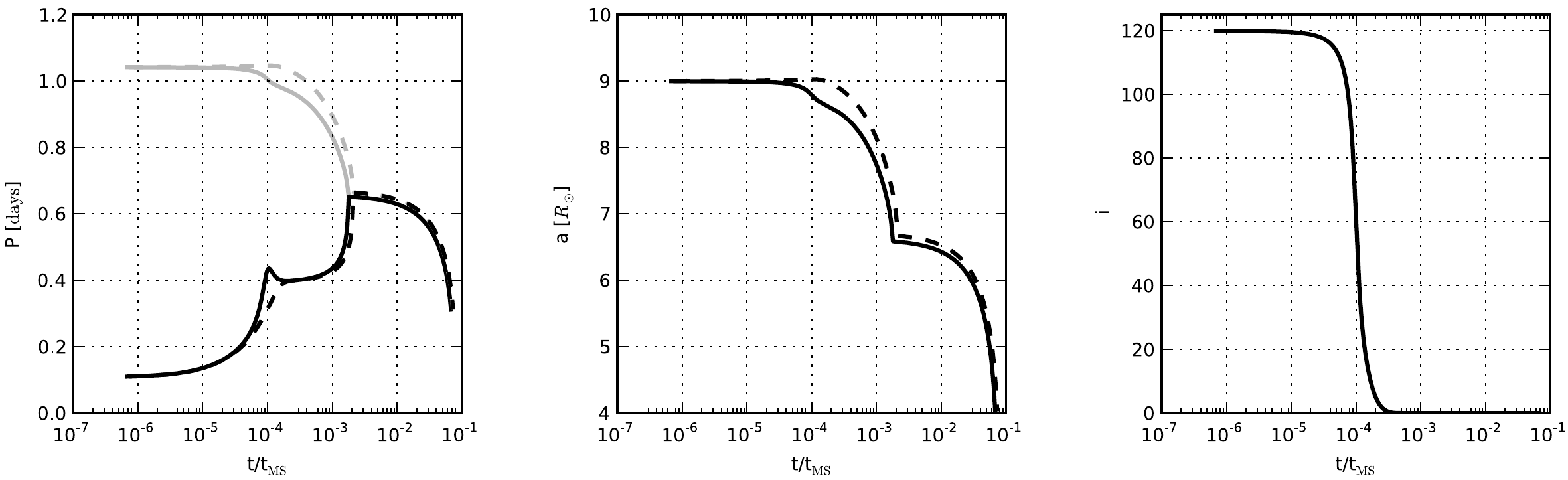}
\caption{
Evolution of a misaligned and eccentric BH-LMXB under the effect of tides and magnetic braking.
The binary contains a black-hole of mass $8~M_\odot$
and a one solar-mass star. 
Initial orbital parameters are
$a=9~R_\odot$, $e=0.5$, $i=120^\circ$.
The left panel show the evolution
of the rotation period of the star
(black solid line)
and of the orbital period (grey solid line).
The middle panel shows the evolution
of the orbital separation (solid black line).
The right panel shows the evolution of the eccentricity (solid black line).
The 
dashed lines in all of the three panels correspond
to the evolution
of a BH-LMXB
with the same initial condition
for ($\omega_\star$, $a$, $e$),
but with no misalignment between the
spin and the orbit.
}
\label{fig:BH-LMXBmisaligned}
\end{figure*}
This is due to the fact that $\omega_\star \cos{i} < \omega_\mr{orb}$.
In the retrograde case,
or in any case when the previous condition is satisfied,
there is a first phase in which 
tides and magnetic braking are simultaneously at work.
In the coplanar case, instead,
the first phase of the evolution
is dominated either by tides
or by magnetic braking,
that act in opposite directions.

In case magnetic braking is not present,
the star would align,
but on a longer timescale.
This is due to the fact that magnetic braking 
is much more efficient than tides
in spinning down the initially high-spinning star.
Once the star has been spun-down,
the spin rapidly aligns.
For the binary configuration we've just showed
the ratio between the time at which 
the spin aligns in
the magnetic braking case
and in the tides-only case
is $\approx 0.6 $.\\

\subsubsection{Population study}
As a more general diagnostic 
of the discrepancy between the coupled and the non-coupled method,
we now do a population study,
investigating how 
both the orbital separation 
at the time the binary circularizes
and the time it takes
to reach circularization,
change over the population.
In the non coupled method, the
two quantities just mentioned are given by
$a_\mr{circ}=a(1-e^2)$ and 
$\tau_\mr{circ}=e/\dot{e}$.\\
{\indent The initial values $e_0$ of the eccentricity }
are taken from a grid 
which spans the interval $[0.001, 0.901]$.
For each value of the eccentricity,
we draw uniform values $a_0$ 
for the orbital separation
in the interval $[a_\mr{min}(e), a_\mr{max}(e)]$.
$a_\mr{min}(e)=a_\mr{RLO}/(1-e)$ is the orbital separation
at which the system undergoes RLO at periastro;
$a_\mr{max}(e)$ is the maximal orbital separation
such that the system undergoes RLO
within the MS-lifetime.
We draw uniform values $\omega_{\star,0}$ for the stellar spin
in the interval $[0, \omega_\mr{break}]$,
and uniform values $i_0$ in $[0^ \circ, 180^ \circ]$.

For each initial condition
($a_0, e_0, \omega_{\star,0}, i_0$),
we integrate the tidal equations
coupled with magnetic braking
and we study the properties of $1000$ solutions
which undergo RLO within the MS-lifetime.
We calculate the ratio
between the value in the coupled method and the value in the non-coupled method
of the previously mentioned variables
as a function of the eccentricity.
We show the results in figure \ref{fig:RatioAcircBH+star} and figure \ref{fig:RatioTimescalesBH+star}.

For each value of the eccentricity there is a range of values for
$t_\mr{circ}/\tau_\mr{circ}$. The spread in the values at a fixed eccentricity is 
mainly caused by different initial spin frequencies.
This was already seen in figure \ref{fig:bhstar},
where 
taking either a low-spinning star
or a high spinning star,
under-estimates or over-estimates
$t_\mr{circ}$.

\begin{figure*}
\begin{minipage}{\columnwidth}
\centering
\includegraphics[width=\textwidth]{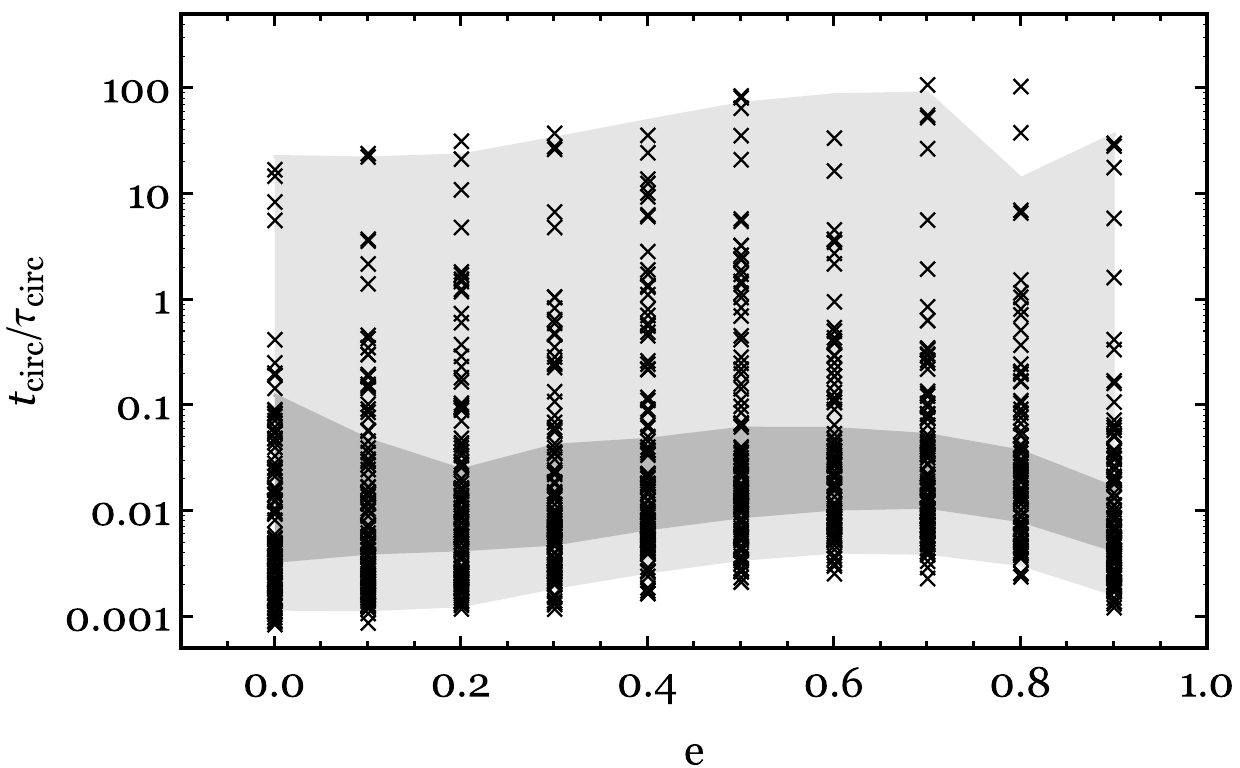}
\caption{Ratio between the time it takes to reach circularization
in the coupled and non-coupled method for 1000 black-hole low-mass X-ray binaries
undergoing RLO within the MS-lifetime,
when taking $\tau_\mr{circ}=e/\dot{e}$.
The light grey shaded
area is for a fixed initial orbital separation,
and the darker grey area 
is for a fixed initial stellar spin.
}
\label{fig:RatioTimescalesBH+star}
\end{minipage}
\hspace{0.3cm}
\begin{minipage}{\columnwidth}
\includegraphics[width=\textwidth]{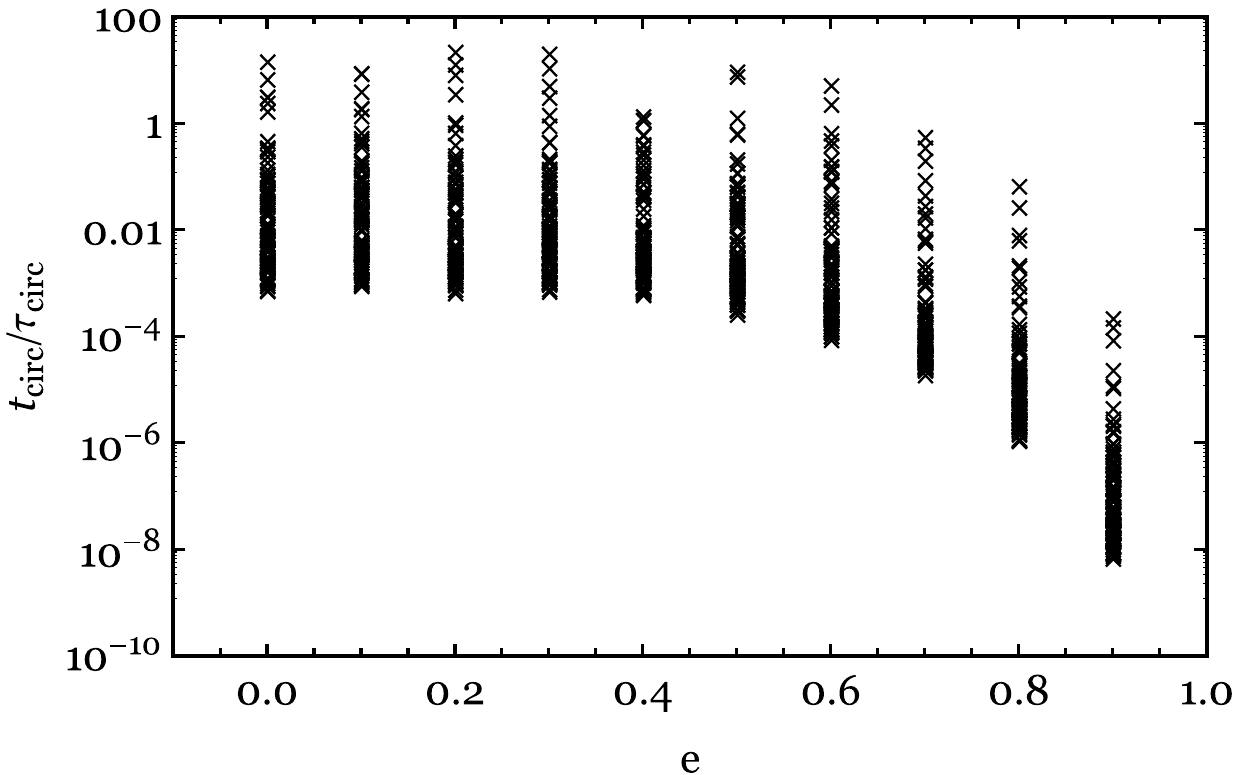}
\caption{Ratio between the time it takes to reach circularization
in the coupled and non-coupled method for 1000 black-hole low-mass X-ray binaries
undergoing RLO within the MS-lifetime,
when taking $\tau_\mr{circ}$
as in eq. \ref{eq:tau_circ}.}
\label{fig:taucircHurley}
\end{minipage}
\end{figure*}

To show how the spread in the ratio $t_\mr{circ}/\tau_\mr{circ}$
is reduced when fixing the initial spin of the star,
we show in figure \ref{fig:RatioTimescalesBH+star}
the ratio $t_\mr{circ}/\tau_\mr{circ}$
when the initial spin
is chosen to be $\omega_\star=1/2~\omega_\mr{break}$
(dark grey shaded area).
When we fix instead the orbital separation at 
birth
(as an example, 
we take the average value
between $a_\mr{min}(e)$
and $a_\mr{max}(e)$),
the spread is conserved
(light grey shaded area).

We also calculate the ratio
$t_\mr{circ}/\tau_\mr{circ}$
when taking for $\tau_\mr{circ}$
the expression in equation \ref{eq:tau_circ},
see figure \ref{fig:taucircHurley}.
The decrease in the ratio is
caused by the dependence
of $\tau_\mr{circ}$
on the orbital separation,
which increases with increasing eccentricities,
and $\tau_\mr{circ}$ increases accordingly.

Concerning the ratio 
between the orbital separation at $e=0$ in our integration
($a_{e=0}$) and
$a_\mr{circ}=a (1-e^2)$,
it is typically less than $1$.
This is due to the fact
that as soon as magnetic braking has spun-down
the star sufficiently,
tides start removing angular momentum from the orbit,
before synchronization is achieved.
This effect is neglected in the non-coupled method.
The decrease 
of the ratio with the eccentricity,
is a consequence
of the $a_\mr{circ}$ dependence
on ($1-e^2$).

We also calculate
the ratio between the time it takes to reach the RLO
configuration
in our integration ($t_\mr{RLO}$)
and the estimated value for it
in the non-coupled method ($\tau_\mr{RLO}$).
In the non-coupled method,
we calculate $\tau_\mr{RLO}$
in two ways.
One as $\tau_\mr{circ}+t_\mr{VZ}$,
the other as $\tau_\mr{VZ}$.
In the first way we integrate the equation 
\ref{eq:MBonTheBinary}
taking as initial condition
$a_\mr{circ}=a  (1-e^2)$;
this gives $t_\mr{VZ}$.
We then calculate 
$\tau_\mr{RLO}$ as $\tau_\mr{circ}+t_\mr{VZ}$,
where $\tau_\mr{circ}=e/\dot{e}$.
The other way relies
on assuming instantaneous circularization
and taking $\tau_\mr{RLO}=\tau_\mr{VZ}$,
where $\tau_\mr{VZ}$
is $a/\dot{a}$
with $\dot{a}$
as in equation \ref{eq:MBonTheBinary},
and again
we use $a_\mr{circ}=a  (1-e^2)$
when calculating $\tau_\mr{VZ}$.
This way of estimating $\tau_\mr{RLO}$
has been mostly used
in previous BPS works
on the evolution
of binaries hosting a black-hole or a neutron star.
We show the respective outcomes
in figures \ref{fig:TimeForReachingRLO}
and \ref{fig:TimeForReachingRLO_VZ}. The decrease in the ratio
which appears in figure \ref{fig:TimeForReachingRLO_VZ}
is due to the fact
that for higher eccentricities,
both $a_\mr{min}$
and $a_\mr{max}$ increase;
$\tau_\mr{VZ}$
increases accordingly.
When choosing $\tau_\mr{RLO}$ as $\tau_\mr{circ}+t_\mr{VZ}$,
the ratio is constrained to be in the interval 
$(0.1-1)$,
so we find this way of calculating
$\tau_\mr{RLO}$ a better estimate
than when $\tau_\mr{RLO}=\tau_\mr{VZ}$.

\begin{figure*}
\begin{minipage}{\columnwidth}
\centering
\includegraphics[width=\columnwidth]{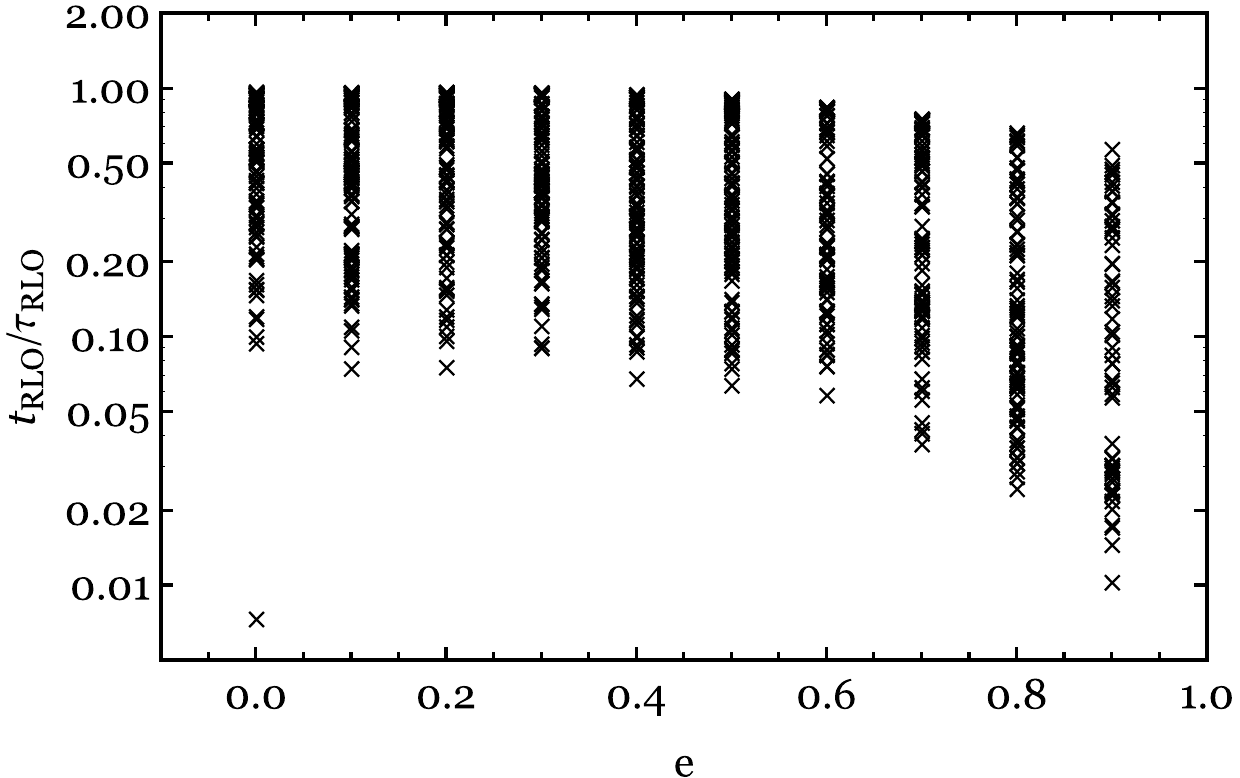}
\caption{Ratio between the time it takes to reach RLO
in the coupled and non-coupled method for 1000 black-hole low-mass X-ray binaries
undergoing RLO within the MS-lifetime.
In the non coupled method,
we take $\tau_\mr{RLO}$ 
as $\tau_\mr{circ}+t_\mr{VZ}$ (see text for details).
}
\label{fig:TimeForReachingRLO}
\end{minipage}
\hspace{0.3cm}
\begin{minipage}{\columnwidth}
\centering
\includegraphics[width=\columnwidth]{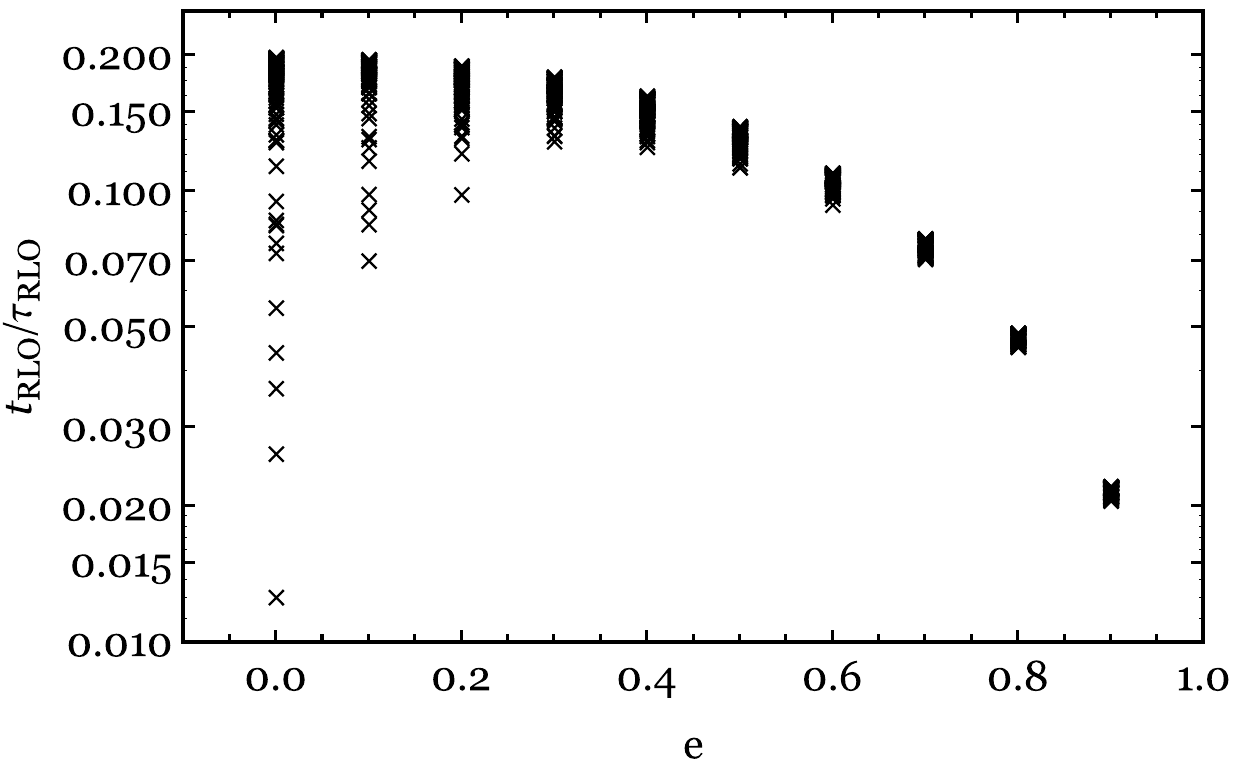}
\caption{Ratio between the time it takes to reach RLO
in the coupled and non-coupled method for 1000 black-hole low-mass X-ray binaries
undergoing RLO within the MS-lifetime.
In the non coupled method,
we take $\tau_\mr{RLO}$ 
as $\tau_\mr{VZ}$ (see text for details).
}
\label{fig:TimeForReachingRLO_VZ}
\end{minipage}
\end{figure*}

\begin{figure}
\centering
\includegraphics[width=\columnwidth]{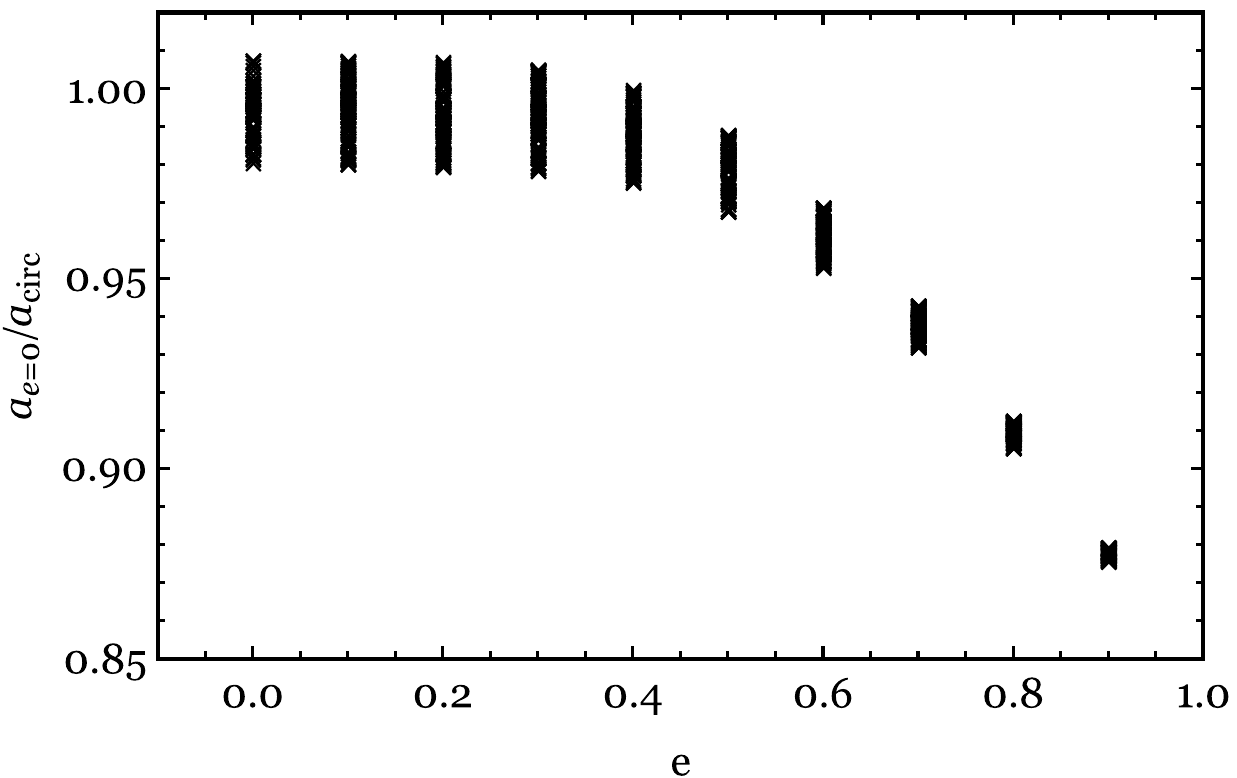}
\caption{Ratio between the orbital separation at $e=0$
and $a_\mr{circ}$
for 1000 black-hole low-mass X-ray binaries
undergoing RLO within the MS-lifetime.}
\label{fig:RatioAcircBH+star}
\end{figure}

\subsection{Planetary system}
We follow the evolution of a planetary
system as well,
composed of a 
Sun-like star
and a Hot Jupiter.

\subsubsection{An illustrative example of the evolution in the coplanar case}
We show an illustration of the evolution until the planet fills
its Roche-lobe
(at $a_\mr{RLO}\approx 2~\Rsun$)
in figure \ref{fig:planet+staraligned}.
The initial configuration is
$a=4~R_\odot$, $e=0.2$, $i=0$, and
we again consider the two cases
of a high-spinning star ($0.9~\omega_\mr{break}$)
and a low-spinning star ($10^{-5}~\omega_\mr{break}$).

In the high-spinning case,
magnetic braking
pushes $\omega_\star$ 
below corotation,
and Hut's stability condition is recovered
within a short timescale.
Below corotation,
tides are too inefficient
for
synchronizing the spin 
within the main-sequence lifetime.
The same result 
was found by BO2009
(see their figure 3).
The difference is that we use a different calibration factor 
causing tides in our model
to be weaker. This is shown expressing the tidal calibration factor
$K/T$
as $\frac{3}{2Q^\prime} \frac{1}{\omega_\mr{orb}} \frac{G M_\star}{R_\star^3}$.
This leads to $K/T\approx 7\times 10^{-9}$
for the chosen initial conditions
and $Q^\prime = 10^6$,
whereas in our model
$K/T\approx 10^{-11}$,
for every value of the initial spin.
The weaker tides
cause magnetic braking to spin down the star
even more significantly
below corotation. 

In the BH-LMXB case,
a condition in which $\dot{\omega}_{\star\mr{,MB}}=\dot{\omega}_{\star\mr{,tid}}$
is reached. This does not happen in the planetary case.
The planetary system typically goes through
one or two main evolutionary phases,
each driven either by magnetic braking or by the tidal torque.
In the high spin case, the first phase is driven
by MB until the spin finds its-self below corotation,
at which point the evolution
is driven by the tidal interaction.
In the low-spin case,
the whole evolution
is driven by tides.
This overall behaviour of the solution
persists in case the calibration factor is the one in BO2009.
The stellar spin does not converge to the quasi-equilibrium
characterized by $\omega_\star=\omega_\mr{eq}$.
In both tidal models,
$\omega_\mr{eq}$
is much smaller than typical
$\omega_\mr{eq}$ in BH-LMXBs
(the corresponding $P_{\star\mr{,eq}}$
is $\approx 21$ days
-see dotted line in the left panel of figure \ref{fig:planet+staraligned}-
and $\approx 3 $ days in BO2009 model for tides).
{{We will further comments on the absence or presence
of the quasi-equilibrium state in section \ref{sec:quasi}.}}
The only difference between the evolution
in our model and in BO2009 one,
is that in the low-spinning case,
the time it takes
for tides to significantly spin-up the star is shorter.

In the high-spin case,
once $\omega_\star$ has been brought by MB
below corotation
(at $\mr{t}\approx 10^{-3}~\mr{t}_\mr{MS}$),
tides start removing angular momentum from the orbit,
the binary shrinks, and the solution approaches
the low-spinning solution,
since the star now spins too slowly 
for magnetic braking to be effective.

In both cases
of a high-spinning star and a low-spinning star,
tides more easily bring changes to the orbit
than to the star,
due to the high ratio $J_\star/J_\mr{orb}$ (see figure \ref{fig:Overall}),
and in neither case
the RLO-configuration
is synchronous,
due to low mass-ratio,
unlike the BH-LMXB case.
\begin{figure*}
\includegraphics[width=\textwidth]{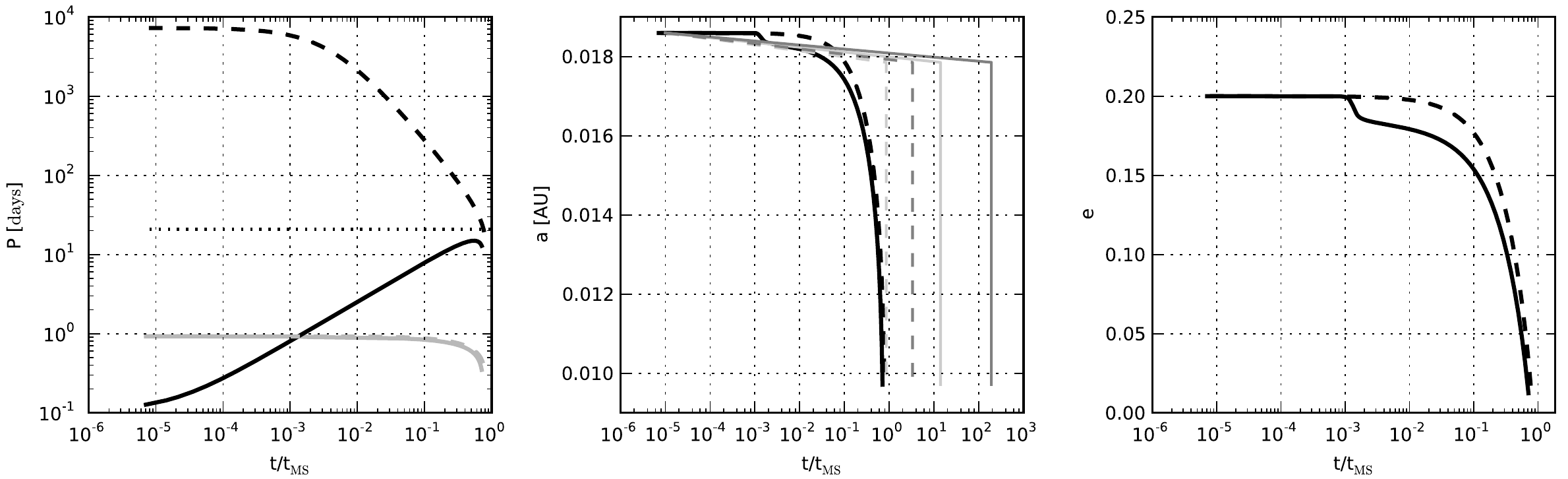}
\caption{Evolution of an eccentric and coplanar planetary system under the effect of tides and magnetic braking.
The system contains a Hot-Jupiter and a sun-like star.
Initial orbital parameters are $a=4~R_\odot$, $e=0.2$, $i=0$.
We consider both a high-spinning star ($\omega_\star=
0.9~\omega_\mr{break}$) and
a low-spinning star ($\omega_\star=10^{-5}~\omega_\mr{break}$).
In all panels,
solid lines correspond to high-spin, dashed lines to low-spin.
In the left panel, the grey lines represent the evolution of the orbital period,
the black lines of the stellar period.
The dotted line in the same panel
represents the value of $\omega_\mr{eq}$ (see
text for details).
In the center panel, 
the two thick lines represent the evolution of the orbital separation in our integration.
In the same panel,
the thin dark-gray lines
represent the evolution of the orbital separation
in the non-coupled method
when taking $\tau_\mr{circ}=\left .\left ({e}/{\dot{e}}\right )\right\vert _\mr{e\approx0, \omega_\star=\omega_\mr{orb}}$;
light-gray lines when taking $\tau_\mr{circ}=e/\dot{e}$.
The right panel shows the evolution of the eccentricity.}
\label{fig:planet+staraligned}
\end{figure*}

In the middle panel of figure \ref{fig:planet+staraligned} we also show
the evolution 
in the non-coupled method,
when taking
$\tau_\mr{circ}=e/\dot{e}$
(thin light-grey lines),
and 
$\tau_\mr{circ}=\left .\left ({e}/{\dot{e}}\right )\right\vert _\mr{e\approx0, \omega_\star=\omega_\mr{orb}}$
(thin dark-grey lines).
Neglecting the spin of the star results
in a different evolution:
the circularization timescale
is overestimated.
At these large values of $J_\star/J_\mr{orb}$,
it is essential to follow 
the coupled evolution
of the orbital and rotational elements.\\

\subsubsection{An illustrative example of the evolution in the misaligned case}
Due to the high ratio $J_\star/J_\mr{orb}$,
we expect tides to be inefficient
in washing away any initial misalignment.
In figure \ref{fig:planet+starmisaligned}, we show the evolution
of the planetary system
for an initial configuration
$a=3~\Rsun$, $e=0$, $i=160^\circ$.
\begin{figure*}
\includegraphics[width=\textwidth]{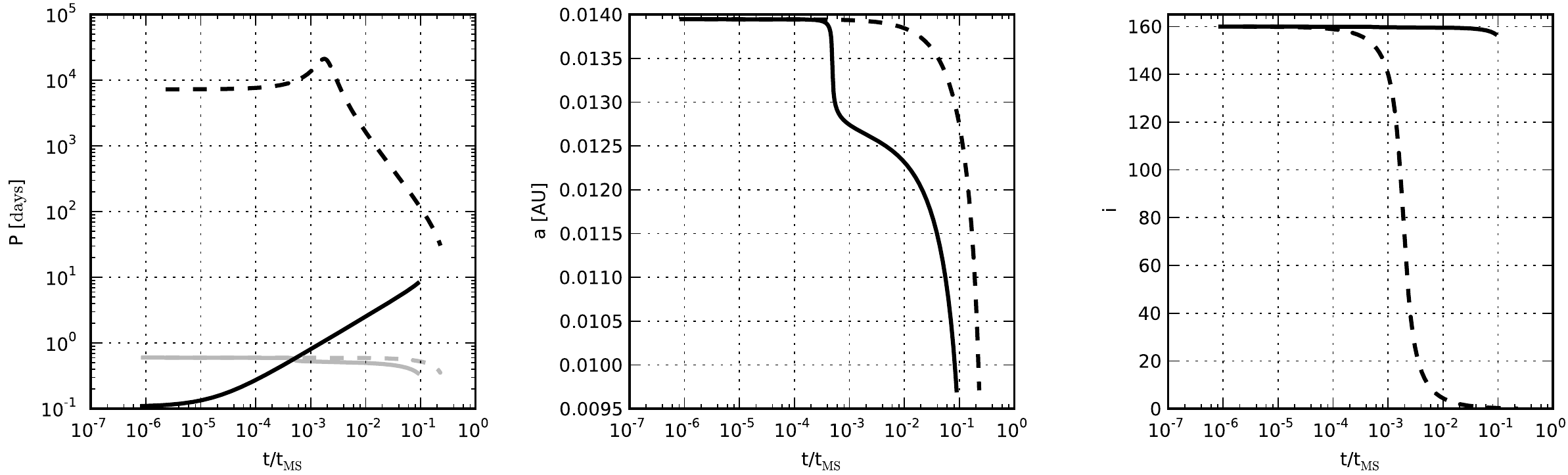}
\caption{
Evolution of an eccentric and misaligned planetary system under the effect of tides and magnetic braking.
The system consists of a Hot-Jupiter and a sun-like star.
Initial orbital parameters are $a=3~R_\odot$, $e=0$, $i=160^\circ$.
We consider both a high-spinning star ($\omega_\star=
0.9~\omega_\mr{break}$) and
a low-spinning star ($\omega_\star=10^{-5}~\omega_\mr{break}$).
In all panels,
solid lines correspond to high-spin, dashed lines to low-spin.
In the left panel, the grey lines represent the evolution of the orbital period,
the black lines of the stellar period.
In the center panel, 
we show the evolution of the orbital separation.
{{The right panel shows the evolution of the misalignment angle}}.}
\label{fig:planet+starmisaligned}
\end{figure*}
The rate of alignment
is larger in the low-spin case,
whereas in the high-spin case,
the star rotational angular momentum 
is too large for being affected by the tidal torque,
and the system reaches RLO
while being in a non-coplanar configuration.
We might wonder what would happen to a retrograde orbit in the absence of magnetic braking.
Taking the semi-major axis decay rate for a misaligned orbit (equation \ref{eq:eq1})
and considering a circular case,
the $i$-dependence
of $\dot{a}$ can be written as:
\begin{equation}
\frac{da}{dt}\propto -\left ( 1+ \frac{\omega_\star |\cos{i}|}{\omega_\mr{orb}}   \right )
\end{equation}
The orbit would undergo an initial decay,
with a larger rate for larger inclinations. 
This is an interesting way of shrinking the orbital separation,
though the initial orbital separation 
has to be already 
comparable to
$a_\mr{RLO}$
for RLO to happen on the MS.
For example, 
for the case of an initial misalignment
of $160^\circ$,
the maximum initial orbital separation
is $\approx 1.5~a_\mr{RLO}$.
The same orbital decay due to
$\omega_\star \cos{i} < \omega_\mr{orb}$
happens in a BH-LMXB.
In this case however,
the system always reaches alignment and synchronization within the MS-lifetime for 
sufficiently tight initial orbits,
since the tidal torque is much larger.

\subsubsection{Population study}
We do a population study for the planetary system too,
drawing the initial conditions in the same way
as we did for the BH-LMXB case (see figures \ref{fig:RatioAcircPlanet+Star} and \ref{fig:RatioTimescalesPlanet+Star}).
In the non-coupled method,
$\tau_\mr{circ}$ is again $e/\dot{e}$.
\begin{figure*}
\begin{minipage}{0.45\linewidth}
\centering
\includegraphics[width=\textwidth]{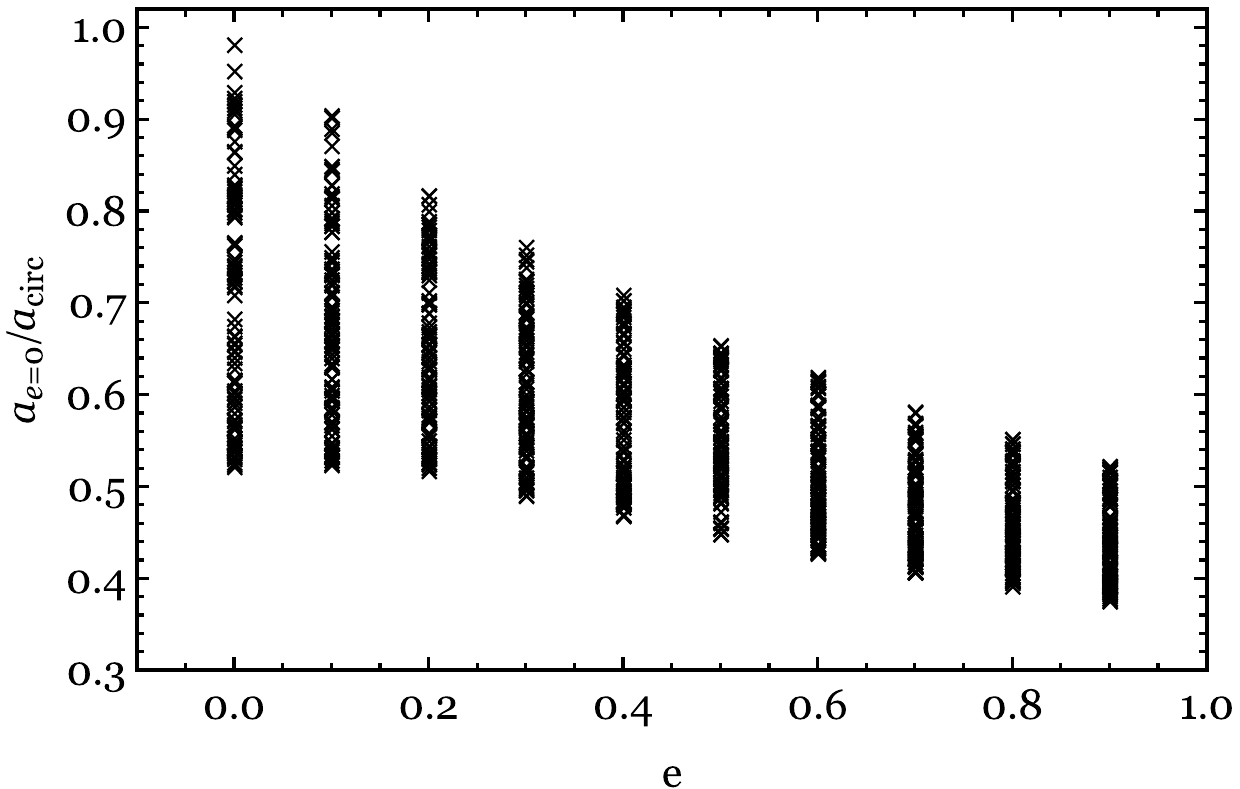}
\caption{Ratio between the orbital separation at $e=0$
and $a_\mr{circ}$ 
for 1000 planetary systems
undergoing RLO within the MS-lifetime.}
\label{fig:RatioAcircPlanet+Star}
\end{minipage}
\hspace{0.3cm}
\begin{minipage}{0.45\linewidth}
\centering
\includegraphics[width=\textwidth]{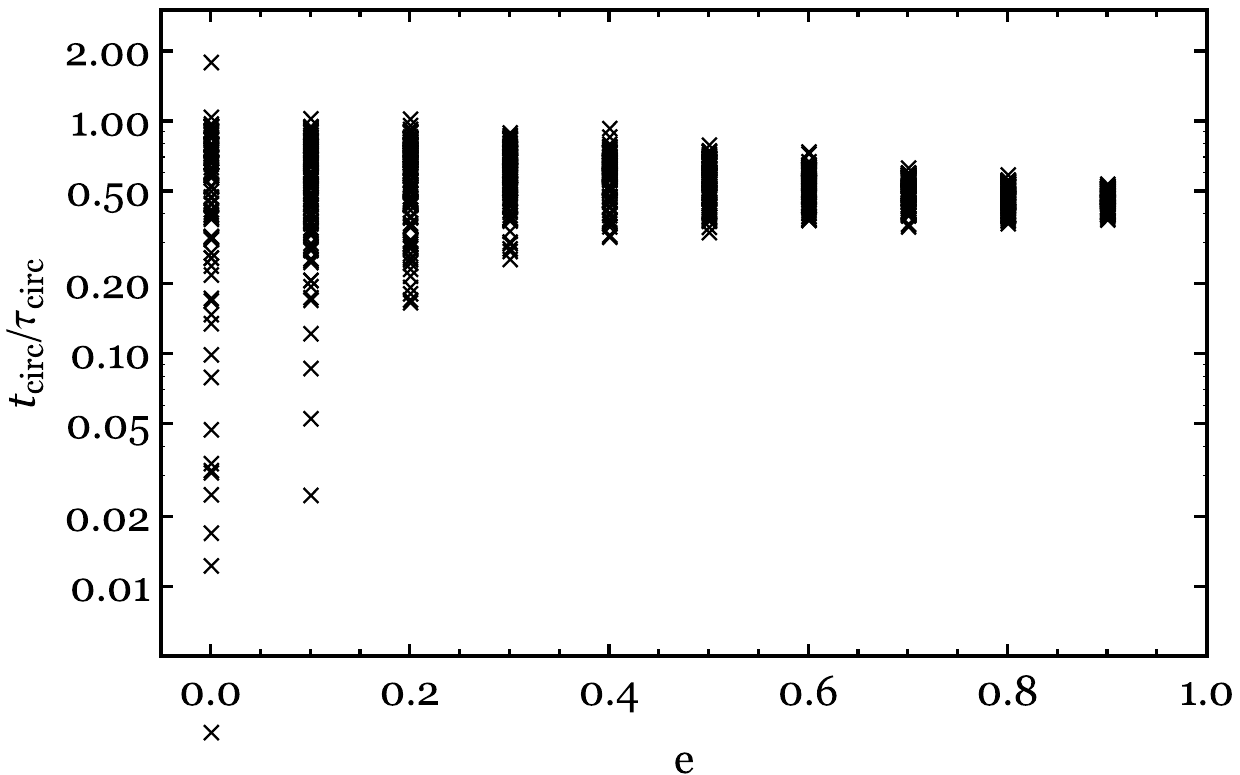}
\caption{Ratio between the time it takes to reach circularization
in the coupled and non-coupled method for 1000 planetary systems
undergoing RLO within the MS-lifetime.
}
\label{fig:RatioTimescalesPlanet+Star}
\end{minipage}
\end{figure*}

Again we see that $a_\mr{circ}$ typically overestimates
the actual orbital separation at $e=0$.
Concerning the ratio of the timescales, 
it spans a range of $\approx$ $[10^{-2}$-$1]$.
This is consistent with the example of figure
\ref{fig:planet+staraligned}.

\section{Wind Braking}
\label{sec:windbraking}
While
most low-mass stars are magnetic,
due to the dynamo processes undergoing in their convective envelopes,
only a fraction of intermediate/high-mass stars show magnetic fields.
This fraction is smaller than about $15\%$ of the total population,
see \citealt{2009ARA&A..47..333D}.
When the massive star is magnetic,
the stellar wind becomes an analogue of the magnetic-wind in a low-mass star,
the only difference being 
that mass loss is non-negligible anymore.
We will take into account the effect of the mass-loss on the binary evolution
adding a mass-loss term both in
the orbital separation rate and in 
the spin-frequency rate. We will show that
neglecting the spin-down rate due to the mass-loss
leads to a different evolution.

\subsection{Tides coupled with wind braking}
\label{sec:wbequations}
Mass loss in a stellar wind removes angular momentum 
from a rotating star.
We assume that the wind is radial and isotropic;
the wind can hence be modeled as a spherical shell decoupling from the star at a certain {\it{decoupling radius}} $r_\mr{d}$.
If the star is non magnetic, 
$r_\mr{d}$ is the radius of the star $R_\star$;
whereas if it is magnetic, $r_\mr{d}$ is the magnetospheric radius $r_\mr{M}$,
i.e.
the radius out to which the material corotates with the star.

Expressing the decoupling radius in terms of the stellar radius 
as $r_\mr{d}=\gamma R_\star$, the rate of angular momentum lost is then:
\begin{equation}
\frac{d\vec{J}_\star}{dt}={{\frac{2}{3}\dot{M_\star}\omega_\star \gamma^2 R_\star^2\vec{e}_\omega}}  
\end{equation}
where $\omega_\star$ is the rotational frequency of the star.
This expression coincides with the 
the well-known prescription
for the angular-momentum loss in a stellar wind
by \citealt{1967ApJ...148..217W},
when parametrizing the Alfven radius
in terms of the stellar radius.
It is valid 
both for a stellar wind
decoupling at the stellar surface ($r_d=R_\star$),
and for a wind which is forced to 
corotate out to $r_d$
by a purely radial magnetic field.
For a different 
field geometry 
this expression becomes:
\begin{equation}
\frac{d\vec{J}_\star}{dt}=\frac{2}{3}\dot{M_\star}  \omega_\star R_\star^2  \gamma^n    \vec{e}_\omega  
\label{eq:JdotNonRadial}
\end{equation}
recovering the purely-radial field configuration
for $n=2$
(\citealt{1988ApJ...333..236K}).

We assume that the wind
decouples from the binary at $r_\mr{d}$,
without further interaction with the binary components. This is the so-called {\emph{fast-wind}} approximation,
motivated by the fact that typical wind speeds are larger than typical orbital speeds.
In a neutron-star high-mass X-ray binary 
containing a NS of $1.4~\Msun$ and a $15~\Msun$ star,
$v_\mr{orb}< 600$ km/s
for $a>a_\mr{RLO}$.
Taking as wind-velocity $v_\mr{wind}$
the escape-velocity from the star,
we obtain
$v_\mr{wind}\approx 1000$ km/s.

To obtain the spin-down rate,
we use the mass-radius relation for a ZAMS star,
and we assume that
the radius of gyration does not change during the evolution on the MS:
\begin{equation}
\label{eq:windbraking}
\frac{d\omega_\star}{dt}=-\frac{\omega_\star}{M_\star}\dot{M}_\star-\frac{2\omega_\star}{R_\star}\frac{dR_\star}{dM_\star}\dot{M_\star}+{2\over 3} {\dot{M}_\star {r_d}^2 \omega_\star \over k^2 M_\star {R^2_\star}}
\end{equation}
{\indent{If the star is in a binary,
the wind won't take away rotational angular momentum only,
but orbital angular momentum as well. This effect adds up 
to the tidal effect on the orbital separation
as a non-negative term $-a{\dot{M}_\star}/{M}$
in the orbital separation rate.}}

The full evolutionary equations 
for an eccentric and misaligned
binary system
under the coupled effect of tides
and a massive stellar wind 
{{are equations \ref{eq:eq1}, \ref{eq:eq2}, \ref{eq:eq3}, \ref{eq:eq4},
to which we add the term in equation \ref{eq:windbraking}
and the term:}}
\begin{equation}
\label{eq:massive1}
\frac{da}{dt}  = -\frac{\dot{M}_\star}{M}a 
\end{equation}
and we use the 
calibration factor $(K/T)_\mr{r}$
for a radiative envelope.

The mass-loss
for a sun-like star
on the MS
is negligible,
thus we can neglect the term in eq. \ref{eq:massive1}.
The mass-loss effect on
the spin-down rate
is reduced to the third term 
in equation
\ref{eq:windbraking}.
Even if the mass-loss is small,
this term can be significant,
when the decoupling radius is large
($r_\mr{d}\approx 20~\Rsun$ for the Sun).
The magnetic-braking law \ref{eq:MB}
is empirical,
and it 
can be recovered 
through MHD-theory coupled
with theoretical studies of the dynamo process.
\citealt{1988ApJ...333..236K} 
studied the angular momentum loss
in low-mass stars
and
showed 
that Skumanich's law
is recovered
taking $n=3/2$
in equation
\ref{eq:JdotNonRadial}
and neglecting the change in mass and radius
of the star.

The mass-loss in a fast isotropic wind
always widens the orbit
of a binary formed by point-like components,
due to the decrease in binding energy.
However, if the mass-losing star
suffers from tidal deformation,
the tides-induced torque can prevent the widening
thanks to the redistribution of angular momentum.
When $\omega_\star < \omega_\mr{eq,tid}$, where 
\begin{equation}
\label{eq:omegaeqtid}
{\omega}_{\rm eq,tid} = f_2 \left( e^2 \right) {\omega}_{\rm orb} 
\left[ \frac{1}{f_5 \left( e^2 \right) \left( 1 - e^2 \right)^{3/2}} \right] \:,
\end{equation}
the tidal torque term in
the spin rate is positive (see equation \ref{eq:eq3}).
This means that tides bring angular momentum from the orbit
to the star,
counteracting the effect of the mass loss.
We note that $\omega_\mr{eq,tid}$
is alternatively referred to as
\emph{pseudo-synchronization} frequency,
the synchronization
frequency on an eccentric orbit 
{{(see \citealt{1981A&A....99..126H}).
This is an {\emph{instantaneous}}-equilibrium
spin frequency,
since orbital properties are still evolving. 
When the orbit is circular, this pseudo-synchronization frequency
coincides with the orbital frequency.}}

In the next paragraph,
we will estimate for which value of 
$\gamma$
the effect of the tidal torque
in decreasing the orbital separation is stronger than the effect
of the orbital angular momentum loss
in increasing it.\\

A few of the previous works
on the evolution
of HMXBs
have integrated the tidal
equations
coupled with the angular momentum loss
in the wind
(see for example
\citealt{2012ApJ...747..111W}).
What we do differently,
is to include in our set of equations
the misalignment
between the spin of the orbit
and the stellar spin,
and to parametrize the angular momentum loss
in the wind in terms of $\gamma$,
following
the orbital evolution
in a semi-analytical way
and highlighting 
some interesting outcomes
of the coupling between tides and a massive stellar wind.

\subsection{Results}
\subsubsection{Timescale considerations}
Let's assume we have a synchronized, circular and coplanar orbit.
The orbital separation
changes due to the loss of orbital
angular momentum (term 
of equation {\ref{eq:massive1}}).
It also changes due the tidal
redistribution of angular momentum.
The star loses rotational angular momentum in the wind,
and the tidal torque
counteracts this effect.
The effect of the tidal torque on the orbital separation
can then be found from $\dot{J}_\mr{orb}=\dot{J}_\star$,
which gives the orbital separation decay rate:
\begin{equation}
\left .\frac{da}{dt}\right\vert _\mr{tid}=-\frac{4}{3} \frac{|\dot{M}_\star | \gamma^2 R_\star^2 M}{M_\mr{comp} M_\star a}
\end{equation}
Taking the ratio between $\dot{a}|_\mr{ML}$ and $\dot{a}|_\mr{tid}$, we get:
\begin{equation}
\frac{\left .\frac{da}{dt}\right\vert _\mr{tid}}{\left .\frac{da}{dt}\right\vert _\mr{ML}}=
\frac{4}{3} \left (\frac{\gamma R_\star}{a_2}\right )^2 q
\end{equation}
where $a_2$ is the distance 
of the star from the binary centre-of-mass, $a_2=(M_\mr{comp}/M)a$.
The effect of the loss of $J_\mr{orb}$
is larger
when the star is further away from the centre-of-mass (i.e. the larger $a_2$ is).
Whereas the effect of the tidal torque
is larger the larger the decoupling radius is
(due to $J_\star$-loss being larger),
and/or the larger is the mass-ratio
(due to the tidal torque dependence on q).

We define $\gamma_\mr{min}$
as $\gamma$
such that the condition $\left .\frac{da}{dt}\right\vert _\mr{tid}= \left .\frac{da}{dt}\right\vert _\mr{ML}$
is met. The tidal torque is more
effective than the orbital angular momentum loss
in changing the orbital separation
for $\gamma>\gamma_\mr{min}$:
\begin{equation}
\gamma_\mr{min}(q, \eta)=\sqrt{\frac{3}{4 q}} \frac{q}{1+q} \frac{\eta}{f(1/q)}
\end{equation}
where we parametrized the orbital separation
in terms of the RLO-separation,
$a=\eta a_\mr{RLO}$.

For a NS-HMXB 
composed of a $1.4~\Msun\mr{NS}$ and a $15~\Msun$ companion \footnote{
There are at least 3 known NS-HMXBs with similar component masses,
see \citealt{1995A&A...303..497V}, \citealt{2005AIPC..797..623V}, \citealt{2005A&A...441..685V}.},
the mass ratio is $q\approx0.09$,
and taking $\eta= 2$, we find 
$\gamma_\mr{min}\approx 0.83$.
For the illustrative integration
showed in figure \ref{fig:massiveevo},
$\eta= 1.5$,
and the corresponding $\gamma_\mr{min}$ is $\approx 0.6$.
This means that
{\emph{wind-braking}} \footnote
{We name "wind braking" (WB)
the shrinking of the semi-major axis 
due to the tidal counteracting effect
on the loss of rotational angular momentum,
in analogy with magnetic braking.}
wins over the mass-loss effect for every value of gamma.
This remains valid when the companion mass
spans a range of
$[10-50]~M_\odot$ in mass.

For a BH-HMXB
composed of a $8~\Msun\mr{BH}$ and a $15~\Msun$ companion,
$q\approx 0.53$.
We calculate $\gamma_\mr{min}(q, 2)\approx 1.9$
and $\gamma_\mr{min}(q, 1.5)\approx1.6$.
The latter $\gamma_\mr{min}$
corresponds
to the initial conditions
of figure \ref{fig:BHHMXB}.
For smaller $\gamma$,
we expect the orbital separation to
grow in time
(see dotted line
in figure \ref{fig:BHHMXB}).
As another example we calculate the minimum gamma
for the BH-HMXB Cygnus X-1.
Its component masses are
$M_\star\approx 19~\Msun$
and $M_\mr{comp}\approx 15~\Msun$
(\citealt{2011ApJ...742...84O}).
The correspondent $\gamma_\mr{min}$
for $\eta=2$ is $\approx 2$.

In case wind braking is effective in shrinking the orbit,
we wonder for which values of mass-loss in the wind
and of $\gamma$,
the binary is shrunk down to RLO
within the MS-lifetime.
The spin-down timescale due to the angular momentum loss in the wind is:
\begin{equation}
\label{eq:WBtimescale}
\frac{1}{\tau_\mr{SD}}=\frac{|\dot{\omega}_{\star |{\mr{SD}}}|}{\omega_\star}
\end{equation}
where $\dot{\omega}_{\star |{\mr{SD}}}$
is as in equation \ref{eq:windbraking}.
The spin-down timescale
is then
a function of the mass-loss and of the decoupling radius. If the star is magnetic,
we use the fact that the decoupling radius can 
be defined as the distance from the star
at which the ram pressure of the flow
and the magnetic pressure are balancing each other.
\citealt{2006MNRAS.366.1415J}
show that:
\begin{equation}
r_\mr{d}\sim B_0^{1/2} R_\star^{13/8} |\dot{M_\star}|^{-1/4} (G M_\star)^{-1/8}
\label{eq:magradius}
\end{equation}
where $B_0$ is the surface stellar magnetic field.
We use this expression for $r_\mr{d}$
in order to rewrite the spin-down timescale \ref{eq:WBtimescale} in terms of $B_0$ and of $\dot{M}_\star$.
We are interested in those combinations
of $\dot{M}_\star$ and $B_0$
for which
the star is sufficiently spun-down on the MS.

In figure \ref{fig:WBoverMSandB}
we show
the ratio between the spin-down timescale and the MS-lifetime
for a star of mass $15~\Msun$ 
($\tau_\mr{MS}\approx 1.15\times 10^7$ years)
as a function of the stellar surface magnetic field.
Different curves are presented 
for different wind mass-losses.
Mass-losses are
chosen
as $|\dot{M}_\star|= \left \{ 10^{-9},10^{-8},10^{-7}, 10^{-6} \right \}~\Msun\mr{/yr}$.
\begin{figure}
\centering
\includegraphics[width=\columnwidth]{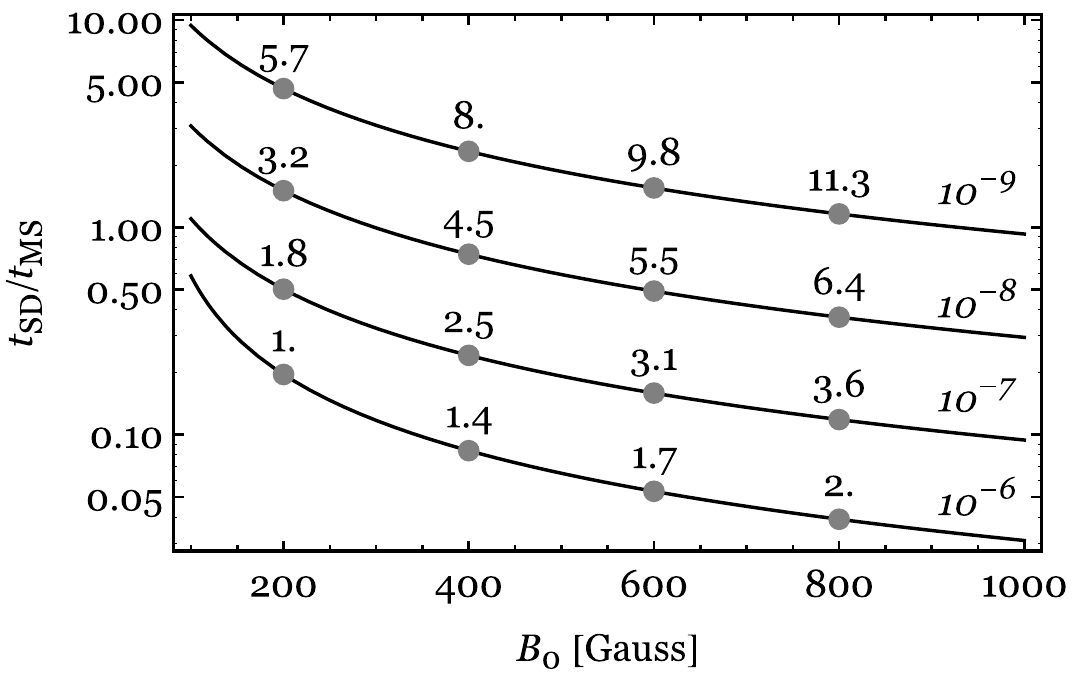}
\caption{
Ratio between the spin-down timescale and the MS-lifetime
for a star of mass $15~\Msun\mr{,}$
as a function of the stellar surface magnetic field
for different mass losses,
$|\dot{M}_\star|= \left \{ 10^{-9},10^{-8},10^{-7}, 10^{-6} \right \}~\Msun\mr{/yr}$.
Values of $\gamma$ as a function
of $B_0$ and $\dot{M}_\star$ are indicated along the curves.
}
\label{fig:WBoverMSandB}
\end{figure}
For larger $B_0$,
the ratio decreases,
since a stronger field
guarantees
the corotation of the field lines
out to large distances from the star,
where the rotational angular momentum carried away by the wind is larger.
For the same magnetic field,
the spin-down timescale is smaller
for larger mass-loss,
since a higher mass-loss
brings away a larger angular momentum.
Along the curves
we also indicate the value of 
$\gamma=r_\mr{d}/R_\star$
for that combination 
of $B_0$ and $\dot{M_\star}$.
For the same field strength at the surface $B_0$,
the decoupling radius is larger 
for a smaller mass loss.
This is due to the fact,
that for a smaller ram pressure (smaller 
$\dot{M}_\star$),
the balancing magnetic pressure ($\sim B_0^2/r_\mr{d}^6$)
is smaller;
if the surface B-field is fixed,
the decoupling radius 
has therefore to be larger.

Typical magnetic fields 
of the subset of magnetic O-B stars 
and of Ap-Bp stars are
of the order of 
hundreds to thousands Gauss (\citealt{2009ARA&A..47..333D}).
We compute their typical main-sequence 
wind mass-loss using the fitting formula
by \citealt{1990A&A...231..134N},
who parametrize the wind-mass loss in terms of
the luminosity, the mass and the radius of the star.
Evolving 
a star with a ZAMS mass of $15~\Msun\mbox{,}$
with the SSE code by 
\citealt{2000MNRAS.315..543H}
embedded in the Astrophysics Multipurpose Software
Environment AMUSE 
(\citealt{2009NewA...14..369P}),
we extract the luminosity and radius of the star.
We find
a wind mass-loss in the range
$10^{-8}-10^{-7}~\Msun\mr{/yr}$ on the MS.

As an example,
for a decoupling radius $\gamma=2$
and a mass-loss $|\dot{M}_\star|=10^{-7}~\Msun\mr{/yr}$,
we obtain an estimate for the required stellar magnetic field
of $\sim 250$ Gauss. For these
values of $B$ and $\dot{M}_\star$,
we expect the binary
to shrink down to RLO.\\

In figure \ref{fig:figure2}
we show the synchronization timescales calculated using eq. \ref{eq:tau_sync}
for a binary formed by a star of mass $15~\Msun$
as a function of the mass-ratio $q$,
both for an orbital separation
$a=a_\mr{RLO}$ and $a=2~ a_\mr{RLO}$.
The assumptions we make
on the decoupling radius and the mass-loss rate are
$\gamma=2$ and $\dot{M}_\star=-10^{-7}~\Msun/\mr{yr}$
which, as can be seen in figure \ref{fig:WBoverMSandB},
allow for a short spin-down timescale 
($\tau_\mr{SD}/\tau_\mr{MS}\approx 0.3$).
For our cases of interest of a NS-HMXB ($q\approx 0.1$)
and of a BH-HMXB ($q\approx 0.5$),
$\tau_\mr{sync}$ is short,
hence we expect the binaries
to rapidly reach the synchronous state.

\begin{figure}
\centering
\includegraphics[width=\columnwidth]{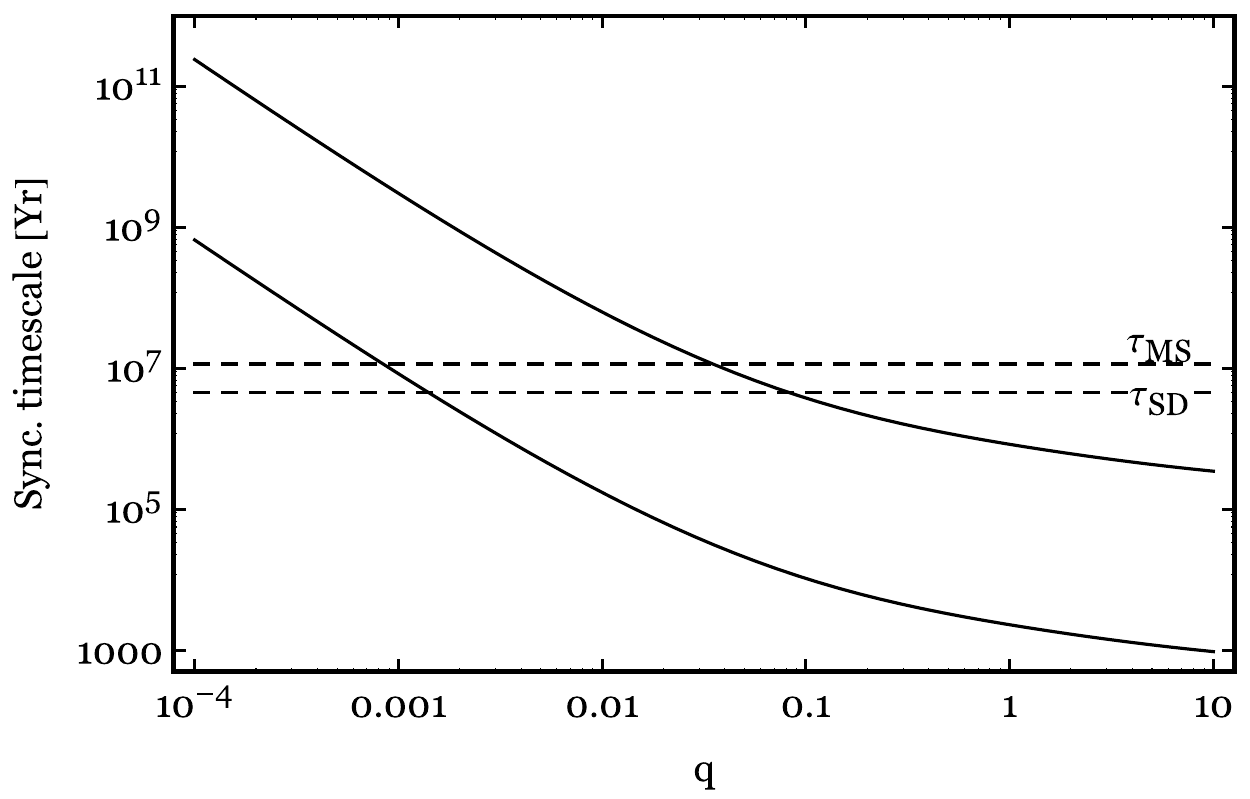}
\caption{
Synchronization timescales
for a binary composed of a $15~\Msun$ star
as a function of the binary mass-ratio $q$, both for an
orbital separation
$a=a_\mr{RLO}$ and $a=2~a_\mr{RLO}$.
The horizontal dashed lines show the spin-down timescale
and the MS-lifetime. The spin-down timescale has 
been computed for $\gamma=2$ and $\dot{M}_\star=-10^{-7}~\Msun/\mr{yr}$.
}
\label{fig:figure2}
\end{figure}

\subsubsection{Wind braking in NS-HMXBs and BH-HMXBs}
\label{sec:NSHMXB}
We integrate the coupled equations
of section
\ref{sec:wbequations}
for a binary composed of
a neutron star of mass $1.4~M_\odot$ and a $15~M_\odot$ star,
with an orbital separation 
$a=14~R_\odot$
and eccentricity $e=0.2$.

We take different combination for the mass-loss rate and the decoupling radius:
$\left \{ |\dot{m}| \mr{,} \gamma\right \}$ = $
\left \{ \left \{ 10^{-7}, 2\right \}\mr{,}\left \{ 10^{-8}, 2\right \}, \left \{ 10^{-7}, 1\right \}\right \}$
(where $\dot{m}$
is the mass-loss rate in solar masses per year).
For each of these three combinations,
we also calculate the evolution
when neglecting the mass-loss terms in the spin-frequency rate
(terms in eq. \ref{eq:windbraking}),
which we call \emph{standard} model.
Concerning the spin frequency, we take a star rotating at $20\%$ of its break-up frequency,
which is a typical lower-limit on the natal rotating speed of a high-mass star
(see \citealt{2009ARA&A..47..333D}).
The associated
ratio $\omega_\star/\omega_\mr{eq,tid}$ is
$\approx 0.7$.
We must point out that 
this is valid in the assumption
that the binary
evolution prior
and during the compact object formation
does not affect the stellar spin.
Anyhow, what is important for our study,
is to compare the initial $\omega_\star$
with $\omega_\mr{eq,tid}$ of equation \ref{eq:omegaeqtid}.
Initial conditions 
with $\omega_\star > \omega_\mr{eq,tid}$
are of no interest,
since the binary widens, preventing RLO.

We show in figure \ref{fig:massiveevo}
the outcome of the evolution. 
\begin{figure}
\includegraphics[width=\columnwidth]{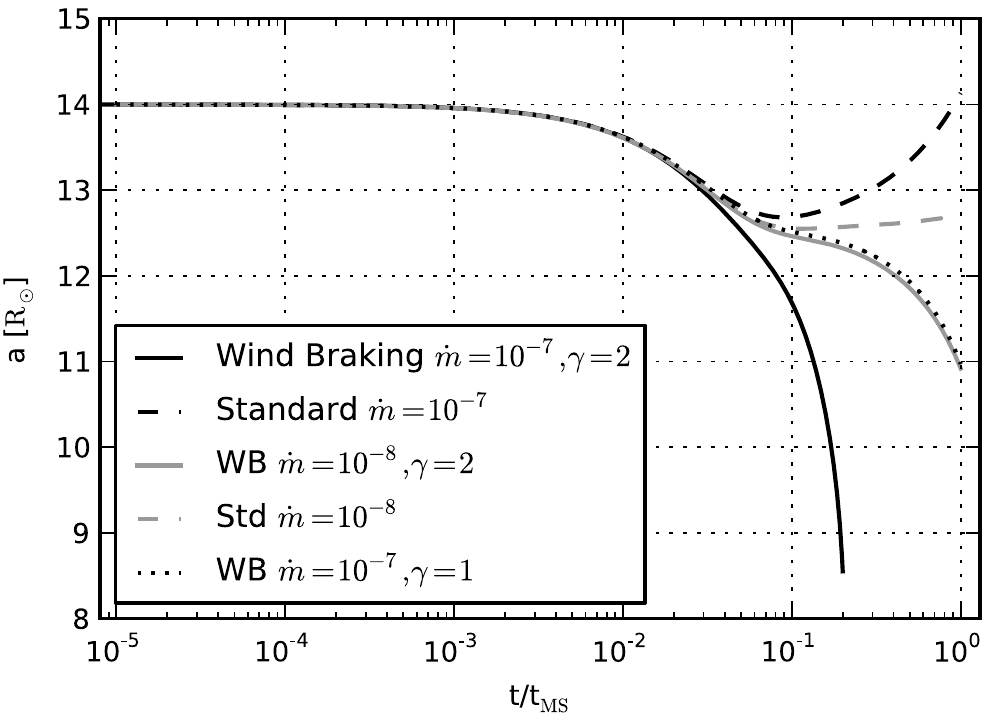}
\caption{Evolution of a NS-HMXB under the effect of tides and a massive stellar wind
(solid lines). Masses of the components are
$M_\star=15~\Msun$ and $M_\mr{comp}=1.4~\Msun$. 
Initial orbital parameters
are $a=14~\Rsun$, $e=0.2$, $\omega_\star=0.2~\omega_\mr{break}$.
We also show the evolution in the standard scenario,
when the mass-loss 
has an effect on the orbital separation only (dashed lines).}
\label{fig:massiveevo}
\end{figure}
The first phase of the evolution
is driven by tides,
until synchronization is achieved
(at $t\approx 6\times10^{-2}~t_\mr{MS}$).
At this moment,
the different prescriptions
on the decoupling radius
and on the mass-loss rate start playing a role.
When dealing with WB,
the evolution is faster for a larger mass-loss.
If the mass-loss is too small,
the binary does not reach RLO
within the MS-lifetime.
If the wind-braking spin-down is not taken
into account,
the orbital separation
increases once the synchronous state
is achieved,
as expected due the positive wind-mass loss term
in the orbital separation rate.
For the wind-braking solutions, once synchronization is achieved,
the tidal torque replenishes
the rotational angular momentum reservoir
of the star,
counteracting the spin-down effect of the wind,
and removing angular momentum from the orbit:
this is exactly what happens in the magnetic braking case.

We also note how
the two solutions
with $\left \{ |\dot{m}| , \gamma\right \}=
\left \{  10^{-7}, 1\right \}$
and $\left \{ |\dot{m}| , \gamma\right \}=
\left \{ 10^{-8}, 2\right \}$
only slightly differ from each other.
This is due to the degeneracy in 
$\dot{m}$ and $\gamma$.
The spin-down timescale is
$\tau_\mr{SD}\approx 2\tau_\mr{MS}$
and $\tau_\mr{SD}\approx 4\tau_\mr{MS}$
for the two choices 
respectively.\\

Solving
the coupled equations
for different values of the initial stellar spin,
we note that if the initial spin is too low,
tides won't manage to synchronize the star on the evolutionary timescale.
What happens is that the orbit keeps shrinking due to
 the tidal torque only,
 and there is no appreciable difference
 between the tides-only solution
 and the WB-solution.
The initial binary configuration $(a, e, \omega_\star)$ which gives rise to 
an evolution of the wind-braking type, satisfies three properties.
The stellar spin is less than $\omega_\mr{eq,tid}$. The configuration allows for tides 
to synchronize the spin within the MS-lifetime. Lastly, it allows the wind to significantly remove angular momentum from the orbit
bringing the system to RLO within the MS-lifetime.

We follow the evolution
under tides and a massive stellar wind
for a BH-HMXB as well.
We take as initial orbital parameters
$a=17~\Rsun$, $e=0.2$,
and a star rotating
at $0.2~\omega_\mr{break}$.
We show in figure \ref{fig:BHHMXB}
the result of the integration.
We note how the solution
with $\left \{ |\dot{m}| , \gamma\right \}=
\left \{  10^{-7}, 1\right \}$
(dotted line)
differs
from the solution
corresponding to the same pair
$\left \{ |\dot{m}| , \gamma\right \}$
for a NH-HMXB.
This is due to the fact that
$\gamma R_\star/a_2$
is smaller for a BH-HMXB
since $a_2$ is larger;
this means that the tidal effect on the orbital 
separation is smaller. This was expected,
since $\gamma=1$ is less than the
minimum value for having wind-braking type
solutions in a BH-HMXB of this type.

\begin{center}
\begin{figure}
\includegraphics[width=\columnwidth]{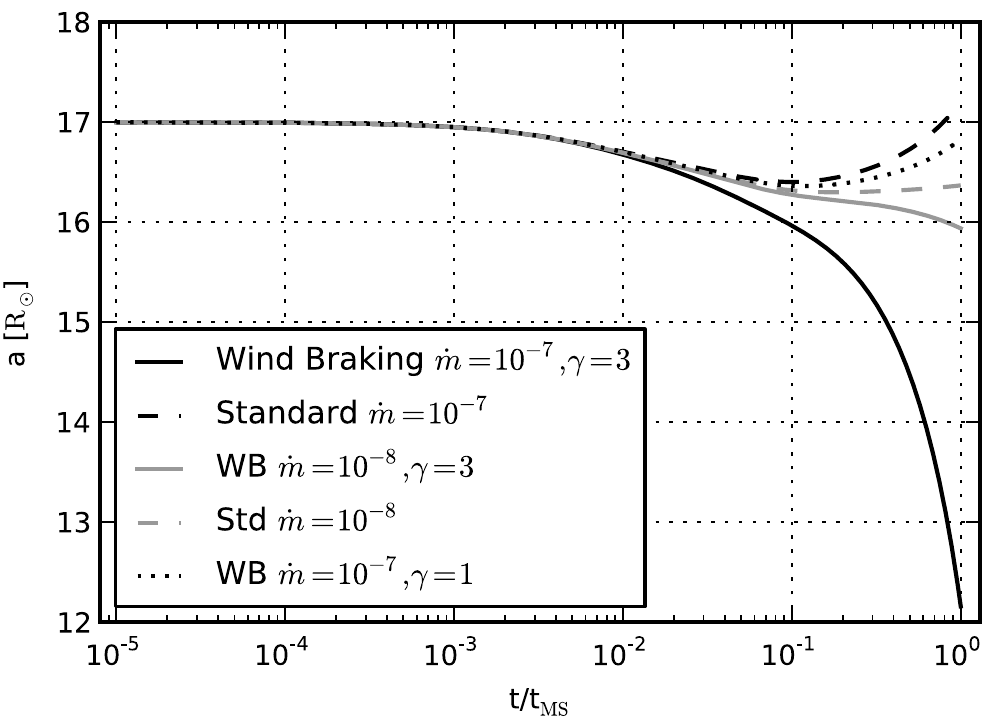}
\caption{Evolution of a BH-HMXB under the effect of tides and a massive stellar wind
(solid lines). 
Masses of the components are
$M_\star=15~\Msun$ and $M_\mr{comp}=8~\Msun$.
Initial orbital parameters
are $a=17~\Rsun$, $e=0.2$, $\omega_\star=0.2~\omega_\mr{break}$.
We also show the evolution in the standard scenario,
when the mass-loss 
has an effect on the orbital separation only (dashed lines).
}
\label{fig:BHHMXB}
\end{figure}
\end{center}

\section{Discussion}
\label{sec:discus}
In this section,
we would like to comment on
the calibration of magnetic braking
and of tides that we chose.
{{We also comment briefly
on how our results on the evolution
of a planetary system might change 
when considering the tides raised
by the star on the planet as well.}}

\begin{enumerate}
\item The empirical Skumanich law
was derived
measuring the 
equatorial
velocity
of G-type MS-stars with rotational velocities between $1-30$ km/s. It is therefore arguable
whether it is still applicable
to fast rotators,
like the ones in compact binaries
where the star,
being synchronized with the orbit,
rotates at velocities of hundreds km/s.
For this reason,
\emph{saturated} magnetic-braking laws have been suggested,
in which the spin-down rate saturates once 
the star-frequency reaches a critical value
(see \citealt{2011ApJS..194...28K},
for reviewing the different MB-prescriptions).
We adopt \citealt{2003ApJ...599..516I}
prescription for MB
in the formulation chosen by
\citealt{2006ApJ...636..985I}:
\begin{equation}
\dot J_\mr{\star,MB} = - K_j \left ( {\frac {R_\star} {R_\odot} } \right )^4
\left\{  \begin{array}{c c}
(\omega_\star/\omega_\odot)^3 \, \, \, \, \, \, \, \ ~{\rm for\ } \omega \le \omega_{\rm x} \\ 
 \omega_\star^{1.3} \omega_{\rm x}^{1.7}/\omega_\odot^3~{\rm for\ } \omega > \omega_{\rm x} 
\end{array} \right. 
\end{equation}
where $\omega_\odot\approx 3 \times 10^{-6}\mr{s}^{-1}$ 
is the angular velocity of the Sun,
$K_{j} = 6\times 10^{30}$ dyn cm,
and the critical angular velocity is
$\omega_x=10~\omega_\odot$.
From this law, the spin-down rate $\dot{\omega}_\star$
is derived.

We integrate the tidal equations coupled 
with the previous MB-prescription
for the BH-LMXB shown in figure \ref{fig:bhstar},
taking a star spinning at $0.9~\omega_\mr{break}$,
a value
which is much larger than the critical angular frequency
$\omega_x$,
thus the spin-down rate scales as $\omega_\star^{1.3}$.
In this prescription
magnetic braking is much less efficient than in the \citealt{1981A&A...100L...7V} prescription, 
hence the binary does not reach RLO within the MS-lifetime
($a\approx 5~a_\mr{RLO}$ at $t=t_\mr{MS}$).
In the first phase of the evolution,
MB spins down the star
while tides lead to a slight widening of the orbit,
unlike what happens in our formulation.
In this case too $\omega_\star$
converges to $\omega_\mr{eq}$,
which corresponds to the value at which
$\dot{\omega}_{\star\mr{,MB}}=\dot{\omega}_\mr{\star,tid}$.
After that, the binary shrinks thanks to the coupling tides-MB,
however much less significantly than 
when using \citealt{1981A&A...100L...7V} prescription for MB.
\item There is a factor $\approx 10^5$
between the timescales 
in our model
and the timescales computed according to BO2009 tidal model.
This is due to the fact that $K/T$
is a
function of both $\omega_\star$ and
$\omega_\mr{orb}$,
through the tidal pumping timescale,
whereas
the calibration factor in 
BO2009,
is a function of $\omega_\mr{orb}$ only.
When $\omega_\star$
and
$\omega_\mr{orb}$ differs strongly
(like in the case when $\omega_\star=0.9~\omega_\mr{break}$)
our tidal calibration factor 
is significantly reduced.
Instead,
when $\omega_\star \approx \omega_\mr{orb}$,
the ratio between the two tidal calibration
factors is $1-2$ for every mass ratio,
leading to similar timescales.

\item We wonder whether our results
are a function
of the tidal calibration we choose.
In the BH-LMXB case,
a different tidal calibration
factor
will affect
the time it takes for reaching 
the circular and synchronized configuration.
However, since this timescale is very short
compared to the MS-lifetime,
independently of the chosen calibration,
we expect no appreciable
difference 
when taking a different tidal calibration factor.
Once synchronization is achieved,
the evolution of the system is well described 
by equation \ref{eq:MBonTheBinary},
that is,
the calibration of tides does not play a role anymore.
Integrating the tidal equations
according to BO2009 model for tides and coupling them with magnetic braking,
we note that when the star is initially high-spinning,
the orbital separation increases initially,
that is in the first phase
of the evolution
tides and MB are both at work;
this will lead to a larger circularized
orbital separation,
but the final outcome of the evolution of a BH-LMXB is not significantly affected.
As a more quantitative
test, we use the maximal initial orbital separation $a_\mr{max}$
such that the BH-LMXB reaches RLO
within the MS-lifetime;
the ratio between $a_\mr{max}$
in our model and $a_\mr{max}$
in BO2009 model of tides,
is $\approx 1$,
for all values of the initial eccentricity.

In the planetary system case,
the situation is different.
A high-spinning star 
is typically spun-down below corotation.
At this point,
MB has a negligible effect,
unlike the tidal interaction, and, if the initial configuration is close enough,
tides bring the system towards RLO.
The stronger tides are,
the faster the evolution towards RLO.
Looking at the maximal initial orbital separation so that 
the planetary system reaches RLO
within the MS-lifetime,
the ratio between $a_\mr{max}$
in our model and $a_\mr{max}$
in BO2009 model of tides,
is $\approx 0.3-0.4$.

{{\item When applying our model to the planetary case,
we only considered the tide raised by the planet on the star,
and not the tide raised by the star on the planet. 
Since the angular momentum stored in the planet
is much less than the orbital angular momentum,
due to the compactness of the planet,
we expect the planet to rapidly synchronize with the orbit.
If tides in the planet are included,
we expect circularization to be reached faster,
due to eccentricity damping effect of the tide in the planet
(see BO2009 and \citealt{1996ApJ...470.1187R}). 
\citealt{2010ApJ...725.1995M} followed the evolution
of a Hot Jupiter in an eccentric and misaligned orbit around its host star,
accounting for both the tide in the star and in the planet.
The stellar tide largely dominates the evolution
of the orbital separation and obliquity.
Instead, both the tide in the planet and in the star
contribute to the damping of the eccentricity,
and depending
on the planet tidal quality factor,
circularization can be achieved without a significant orbital decay.
We expect our main conclusions
on the inadequacy of timescale considerations
to be preserved,
as already predicted by \citealt{2008ApJ...678.1396J},
who accounted for the tide in the planet.
}}
\end{enumerate}

\subsection{On the quasi-equilibrium state}
\label{sec:quasi}
{{The concept of {\emph{quasi-equilibrium state}}
for the spin of the stellar component in a binary,
was already introduced by \citealt{2004ApJ...610..464D},
for the case of a planetary system containing
a star and a Hot Jupiter.
At the quasi-equilibrium,
the rate of angular momentum loss in the stellar wind
is balanced by the rate at which the star gains angular momentum from the orbit
as the planet attempts to spin-up the star.
The authors give an upper limit for the angular momentum loss in the stellar
wind, such that this quasi-equilibrium can be achieved.
We use a similar approach,
and we calculate the maximum spin-down rate due to MB
such that the quasi-equilibrium can be achieved.
In order to find $\omega_\mr{eq}$,
we put $\dot{\omega}=0$ in equation \ref{eq:eq3},
and we take a circular and coplanar orbit.
We obtain:
\begin{equation}
\omega_\mr{eq}=\omega_\mr{orb}-\frac{1}{3} \frac{1}{K/T} \frac{k^2}{q^2} \left ( \frac{a}{R_\star}\right)^6 |\dot{\omega}_\mr{MB}|
\end{equation}
where $a$ and $\omega_\mr{orb}$ are the orbital separation and the orbital frequency
at the quasi-equilibrium state.
This quasi-equilibrium spin value is attainable provided that:
\begin{equation}
\label{eq:omegadotmax}
|\dot{\omega}_\mr{MB}|< \frac{9}{2 Q^\prime} \frac{q^2}{k^2} \frac{G M_\star R_\star^3}{a^6}=\dot{\omega}_\mr{max}
\end{equation}
We note that
we have chosen the calibration factor as in BO2009,
that is we have replaced $K/T$ with $\frac{3}{2Q^\prime} \frac{1}{\omega_\mr{orb}} \frac{G M_\star}{R_\star^3}$,
so that the equation $\dot{\omega}=0$ could be solvable analytically in terms of $\omega_\mr{eq}$.
Condition \ref{eq:omegadotmax} is typically not fulfilled
in a planetary system,
whereas it is fulfilled in a BH-LMXB
(see figure \ref{fig:omegamax}}).
This is consistent with our results
in section \ref{sec:results}.}
\begin{figure}
\centering
\includegraphics[width=\columnwidth]{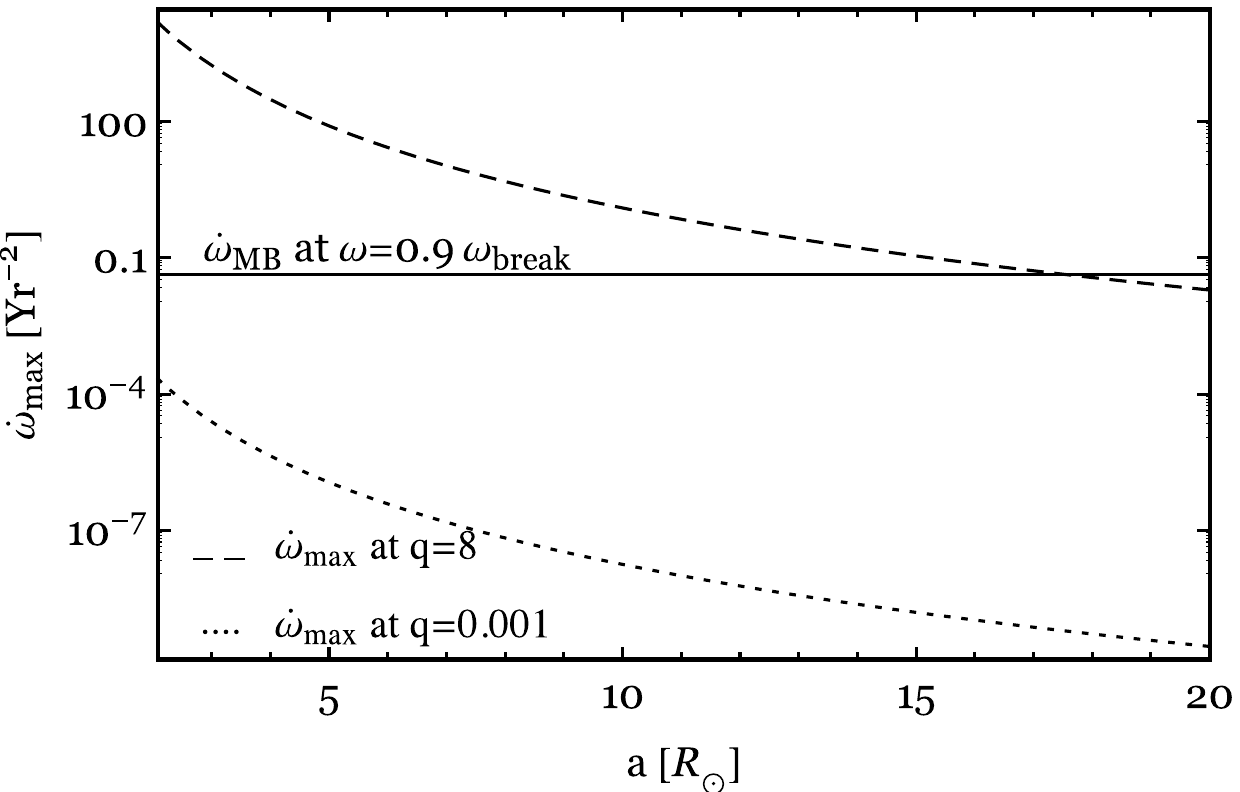}
\caption{Values of $\dot{\omega}_\mr{max}$ such that the system can reach the quasi-equilibrium state,
as a function of the orbital separation, and calculated for a mass ratio $q=8$ and for $q=0.001$.
Horizontal lines correspond to the spin-down rate due to MB.}
\label{fig:omegamax}
\end{figure}

\subsection{An application to the NS-HMXB Circinus X-1}
One way of testing the calibration of tides,
is to look at young
and detached high-mass X-ray binaries,
which still did not have time
to circularize.
In detached systems,
any orbital decay or boost could be directly associated with the tidal interaction
between the two components. Such a study was performed by
\citealt{2008ApJS..174..223B}
for LMC X-4,
a NS-HMXB.
The observed orbital decay of the source,
$P/\dot{P}\approx 10^6$ years,
is consistent
with an orbital decay driven by the tidal torque
when using the calibration
factor in equation \ref{eq:KTrad}.

Another possible example is Circinus X-1,
the youngest known X-ray binary.
An upper limit on its age of $t<4600$ years has been placed
by \citealt{2013ApJ...779..171H},
through
X-ray studies of its natal supernova remnant.
The X-ray emission is thought to be caused by RLO at periastro,
hence this binary is semi-detached,
unlike LMC X-4.
X-ray dip timing
shows the binary is undergoing orbital decay
(\citealt{2003ApJ...595..333P}
and \citealt{2004MNRAS.348..458C}),
with 
$P/\dot{P}$
measured to be $\sim 3000$ yr. 
The current orbital parameters are
$e\approx0.45$ and $P_\mr{orb}\approx16.68$ days.
\citealt{2007MNRAS.374..999J}
constrained the radius
of the companion star
by ensuring
the neutron star
does not go through the companion
at periastro.
The surface gravity gives
in turn the mass.
We use $M_\star=10~M_\odot$
and $R_\star\approx 39~R_\odot$
for the mass and radius of the high-mass star,
and $M_\mr{comp}=1.4~M_\odot$
for the mass of the neutron star.
These initial conditions
allow us
to compute
the current timescales for the change
in the orbital period
$P/\dot{P}=(2/3)a/\dot{a}$,
using the tidal equations in our model.
The only uncertainty
is the spin of the companion.
When $\omega_\star$ is 
smaller than $\omega_\mr{eq,tid}$,
tidal interaction results in a decay of the orbit.
For these initial orbital parameters,
$\omega_\mr{eq,tid}\approx 1.25~\omega_\mr{break}$.
The calculated
timescales are 
$\approx 9\times 10^3$ years and 
$\approx 2\times10^4$ years,
for $\omega_\star$
varying from $0$ and $0.9$ $\omega_\mr{break}$.
{{One possibility is that tides are more efficient than the ones in our model,
by a factor of $\approx 3$
when taking a non-rotating star.
Alternatively,
the observed orbital decay
is either induced by the mass-transfer itself,
or an additional spin-orbit coupling is responsible for it,
as suggested by \citealt{2013ApJ...779..171H}.}}

\subsection{Remarks on our results on the circularization timescale and their consequences}
{{
\begin{enumerate}
\item There is a variety of papers in the literature which base their description
of the evolution of compact binaries on timescale considerations
(see for example, \citealt{1999ApJ...521..723K}, \citealt{2006A&A...454..559Y},
\citealt{1988A&A...191...57P}, \citealt{2009ApJ...691.1611M}).
Since citing all of them does not seem like
a sensible choice,
we rather show an example of
how timescale considerations could fail.
\citealt{2005A&A...439..433J} study a compact binary
formed by a pulsar and a low-mass star.
The observed eccentricity is $e\lesssim 0.005$,
while the initial eccentricity is constrained to be $e\sim 0.7$.
The authors assume an exponential decay of the eccentricity
when dealing with the circularization of the binary.
We instead integrated our equations to solve for
the evolution of the eccentricity of this system.
We obtain $t_\mr{circ}=2\times 10^9$ yr
and $\tau_\mr{circ}=6\times 10^7$ yr
for a low-spinning star,
and $t_\mr{circ}=4\times 10^9$ yr
and $\tau_\mr{circ}=2\times 10^{10}$ yr,
for a high-spinning star. This result
highlights the inadequacy of timescale considerations
when studying the evolution of individual 
systems. 

\item We would also like to point out that some
of the currently used BPS codes 
might treat tidal interaction
using timescale arguments
or assuming instantaneous circularization
(see for example, \citealt{1996A&A...309..179P}).
However, our population models for the progenitors of BH-LMXBs show that the population 
of BH-LMXBs (i.e.
the number of systems that start mass transfer on the MS) might not be
very different, also given the large uncertainties in the calibration
factor.
\end{enumerate}
}}

\section{Conclusions}
\label{sec:conclu}
\begin{enumerate}
\item The evolution of a binary formed by a point-mass and a star 
can be solved relatively easily for
arbitrary mass-loss, 
eccentricity, and inclination
via coupled differential equations.
In this way we can easily investigate how 
the evolution changes as a function
of the binary mass-ratio 
and of the ratio $J_\star/J_\mr{orb}$.
\item When tides are coupled with magnetic braking
the evolution 
of a BH-LMXB
can be separated into two main phases.
The first one is driven either by tides
or by magnetic braking,
depending 
on how fast the star is initially rotating.
In both cases,
the outcome of the first phase of the evolution
is that the spin of the star converges to a quasi-equilibrium value
at which $\left |\dot{\omega}_\star{_\mr{,MB}}\right |=\left | \dot{\omega}_\star{_\mr{,tid}}\right |$: the effect of the tidal torque and of magnetic braking
are balancing each other.
From this moment on,
every piece of angular momentum lost from the star
is subtracted from the orbit.

\item In a planetary system instead,
tides are too weak,
and an initially-high stellar spin is typically brought below corotation.
From this point on,
the evolution coincides with the evolution in the low-spin case.
{{Unlike the BH-LMXB case,
the binary does not reach the quasi-equilibrium state.}}
If the star is initially high-spinning and highly inclined 
with respect to the orbital angular momentum,
the RLO configuration is typically non-coplanar.
\item Models which neglect the coupling between tides and winds do not accurately represent the true evolution of compact binaries. The simple estimate $\tau_\mr{circ}$
is not a good approximation
for the actual change of $e$ over time,
because it does not account for
the change in the orbital separation due to changes in the stellar spin.
Nor for the fact that $\dot{e}$ typically increases during the evolution,
due to decrease of the semi-major axis,
whereas in the estimate $e/\dot{e}$
the orbital separation
$a$ is assumed to be constant.
Furthermore, neglecting the spin leads to
an overestimate
of the semi-major axis at circularization.
It is then essential to consider the coupled evolution
of rotational and orbital elements in order 
to accurately model the evolution
of compact binaries.
This was already showed
previously
by \citealt{2008ApJ...678.1396J}
and \citealt{2009MNRAS.395.2268B},
for the planetary system case.

\item 
We have implemented an easy model
to follow the evolution of a HMXB under
the coupled effect of tides and winds.
For particular choices 
of decoupling radius and mass-loss,
wind braking in a high-mass X-ray binary
behaves as magnetic braking in a low-mass X-ray binary.
The values of decoupling radii and mass-losses which allow for a wind braking-type solution,
are consistent with typical magnetic fields 
and typical mass-losses of high-mass stars.

\end{enumerate}

\section{Acknowledgments}
We thank the anonymous referee whose comments
greatly improved the Paper.
SR is very grateful to Silvia Toonen 
for a careful and critical reading,
which brought significant improvements to the manuscript.
SR is thankful to Adrian Barker
for a useful discussion
on the calibration
of tidal dissipation.
The work of SR was 
supported by the Netherlands Research School for Astronomy (NOVA).

\label{lastpage}


\begin{thebibliography}{}

\bibitem[\protect\citeauthoryear{{Alexander}}{{Alexander}}{1973}]{1973Ap&SS..23..459A}
{Alexander} M.~E.,  1973, \apss, 23, 459

\bibitem[\protect\citeauthoryear{{Barker} \& {Ogilvie}}{{Barker} \&
  {Ogilvie}}{2009}]{2009MNRAS.395.2268B}
{Barker} A.~J.,  {Ogilvie} G.~I.,  2009, \mnras, 395, 2268

\bibitem[\protect\citeauthoryear{{Belczynski}, {Kalogera}, {Rasio}, {Taam},
  {Zezas}, {Bulik}, {Maccarone} \& {Ivanova}}{{Belczynski}
  et~al.}{2008}]{2008ApJS..174..223B}
{Belczynski} K.,  {Kalogera} V.,  {Rasio} F.~A.,  {Taam} R.~E.,  {Zezas} A.,
  {Bulik} T.,  {Maccarone} T.~J.,    {Ivanova} N.,  2008, \apjs, 174, 223

\bibitem[\protect\citeauthoryear{{Church}, {Kim}, {Levan} \& {Davies}}{{Church}
  et~al.}{2012}]{2012MNRAS.425..470C}
{Church} R.~P.,  {Kim} C.,  {Levan} A.~J.,    {Davies} M.~B.,  2012, \mnras,
  425, 470

\bibitem[\protect\citeauthoryear{{Clarkson}, {Charles} \& {Onyett}}{{Clarkson}
  et~al.}{2004}]{2004MNRAS.348..458C}
{Clarkson} W.~I.,  {Charles} P.~A.,    {Onyett} N.,  2004, \mnras, 348, 458

\bibitem[\protect\citeauthoryear{{Darwin}}{{Darwin}}{1879}]{1879Obs.....3...79D}
{Darwin} G.~H.,  1879, The Observatory, 3, 79

\bibitem[\protect\citeauthoryear{{de Mink}, {Langer}, {Izzard}, {Sana} \& {de
  Koter}}{{de Mink} et~al.}{2013}]{2013ApJ...764..166D}
{de Mink} S.~E.,  {Langer} N.,  {Izzard} R.~G.,  {Sana} H.,    {de Koter} A.,
  2013, \apj, 764, 166

\bibitem[\protect\citeauthoryear{{Dobbs-Dixon}, {Lin} \&
  {Mardling}}{{Dobbs-Dixon} et~al.}{2004}]{2004ApJ...610..464D}
{Dobbs-Dixon} I.,  {Lin} D.~N.~C.,    {Mardling} R.~A.,  2004, \apj, 610, 464

\bibitem[\protect\citeauthoryear{{Donati} \& {Landstreet}}{{Donati} \&
  {Landstreet}}{2009}]{2009ARA&A..47..333D}
{Donati} J.-F.,  {Landstreet} J.~D.,  2009, \araa, 47, 333

\bibitem[\protect\citeauthoryear{{Eggleton}}{{Eggleton}}{1983}]{1983ApJ...268..368E}
{Eggleton} P.~P.,  1983, \apj, 268, 368

\bibitem[\protect\citeauthoryear{{Eggleton}, {Kiseleva} \& {Hut}}{{Eggleton}
  et~al.}{1998}]{1998ApJ...499..853E}
{Eggleton} P.~P.,  {Kiseleva} L.~G.,    {Hut} P.,  1998, \apj, 499, 853

\bibitem[\protect\citeauthoryear{{Goldreich} \& {Nicholson}}{{Goldreich} \&
  {Nicholson}}{1977}]{1977Icar...30..301G}
{Goldreich} P.,  {Nicholson} P.~D.,  1977, \icarus, 30, 301

\bibitem[\protect\citeauthoryear{{Heinz}, {Sell}, {Fender}, {Jonker}, {Brandt},
  {Calvelo-Santos}, {Tzioumis}, {Nowak}, {Schulz}, {Wijnands} \& {van der
  Klis}}{{Heinz} et~al.}{2013}]{2013ApJ...779..171H}
{Heinz} S.,  {Sell} P.,  {Fender} R.~P.,  {Jonker} P.~G.,  {Brandt} W.~N.,
  {Calvelo-Santos} D.~E.,  {Tzioumis} A.~K.,  {Nowak} M.~A.,  {Schulz} N.~S.,
  {Wijnands} R.,    {van der Klis} M.,  2013, \apj, 779, 171

\bibitem[\protect\citeauthoryear{{Hurley}, {Pols} \& {Tout}}{{Hurley}
  et~al.}{2000}]{2000MNRAS.315..543H}
{Hurley} J.~R.,  {Pols} O.~R.,    {Tout} C.~A.,  2000, \mnras, 315, 543

\bibitem[\protect\citeauthoryear{{Hurley}, {Tout} \& {Pols}}{{Hurley}
  et~al.}{2002}]{2002MNRAS.329..897H}
{Hurley} J.~R.,  {Tout} C.~A.,    {Pols} O.~R.,  2002, \mnras, 329, 897

\bibitem[\protect\citeauthoryear{{Hut}}{{Hut}}{1980}]{1980A&A....92..167H}
{Hut} P.,  1980, \aap, 92, 167

\bibitem[\protect\citeauthoryear{{Hut}}{{Hut}}{1981}]{1981A&A....99..126H}
{Hut} P.,  1981, \aap, 99, 126

\bibitem[\protect\citeauthoryear{{Ivanov} \& {Papaloizou}}{{Ivanov} \&
  {Papaloizou}}{2004}]{2004MNRAS.353.1161I}
{Ivanov} P.~B.,  {Papaloizou} J.~C.~B.,  2004, \mnras, 353, 1161

\bibitem[\protect\citeauthoryear{{Ivanova} \& {Kalogera}}{{Ivanova} \&
  {Kalogera}}{2006}]{2006ApJ...636..985I}
{Ivanova} N.,  {Kalogera} V.,  2006, \apj, 636, 985

\bibitem[\protect\citeauthoryear{{Ivanova} \& {Taam}}{{Ivanova} \&
  {Taam}}{2003}]{2003ApJ...599..516I}
{Ivanova} N.,  {Taam} R.~E.,  2003, \apj, 599, 516

\bibitem[\protect\citeauthoryear{{Jackson}, {Greenberg} \& {Barnes}}{{Jackson}
  et~al.}{2008}]{2008ApJ...678.1396J}
{Jackson} B.,  {Greenberg} R.,    {Barnes} R.,  2008, \apj, 678, 1396

\bibitem[\protect\citeauthoryear{{Janssen} \& {van Kerkwijk}}{{Janssen} \& {van
  Kerkwijk}}{2005}]{2005A&A...439..433J}
{Janssen} T.,  {van Kerkwijk} M.~H.,  2005, \aap, 439, 433

\bibitem[\protect\citeauthoryear{{Jonker}, {Nelemans} \& {Bassa}}{{Jonker}
  et~al.}{2007}]{2007MNRAS.374..999J}
{Jonker} P.~G.,  {Nelemans} G.,    {Bassa} C.~G.,  2007, \mnras, 374, 999

\bibitem[\protect\citeauthoryear{{Justham}, {Rappaport} \&
  {Podsiadlowski}}{{Justham} et~al.}{2006}]{2006MNRAS.366.1415J}
{Justham} S.,  {Rappaport} S.,    {Podsiadlowski} P.,  2006, \mnras, 366, 1415

\bibitem[\protect\citeauthoryear{{Kalogera}}{{Kalogera}}{1999}]{1999ApJ...521..723K}
{Kalogera} V.,  1999, \apj, 521, 723

\bibitem[\protect\citeauthoryear{{Kawaler}}{{Kawaler}}{1988}]{1988ApJ...333..236K}
{Kawaler} S.~D.,  1988, \apj, 333, 236

\bibitem[\protect\citeauthoryear{{Knigge}, {Baraffe} \& {Patterson}}{{Knigge}
  et~al.}{2011}]{2011ApJS..194...28K}
{Knigge} C.,  {Baraffe} I.,    {Patterson} J.,  2011, \apjs, 194, 28

\bibitem[\protect\citeauthoryear{{Lecar}, {Wheeler} \& {McKee}}{{Lecar}
  et~al.}{1976}]{1976ApJ...205..556L}
{Lecar} M.,  {Wheeler} J.~C.,    {McKee} C.~F.,  1976, \apj, 205, 556

\bibitem[\protect\citeauthoryear{{Ma} \& {Li}}{{Ma} \&
  {Li}}{2009}]{2009ApJ...691.1611M}
{Ma} B.,  {Li} X.-D.,  2009, \apj, 691, 1611

\bibitem[\protect\citeauthoryear{{Matsumura}, {Peale} \& {Rasio}}{{Matsumura}
  et~al.}{2010}]{2010ApJ...725.1995M}
{Matsumura} S.,  {Peale} S.~J.,    {Rasio} F.~A.,  2010, \apj, 725, 1995

\bibitem[\protect\citeauthoryear{{Nieuwenhuijzen} \& {de
  Jager}}{{Nieuwenhuijzen} \& {de Jager}}{1990}]{1990A&A...231..134N}
{Nieuwenhuijzen} H.,  {de Jager} C.,  1990, \aap, 231, 134

\bibitem[\protect\citeauthoryear{{Ogilvie} \& {Lesur}}{{Ogilvie} \&
  {Lesur}}{2012}]{2012MNRAS.422.1975O}
{Ogilvie} G.~I.,  {Lesur} G.,  2012, \mnras, 422, 1975

\bibitem[\protect\citeauthoryear{{Ogilvie} \& {Lin}}{{Ogilvie} \&
  {Lin}}{2007}]{2007ApJ...661.1180O}
{Ogilvie} G.~I.,  {Lin} D.~N.~C.,  2007, \apj, 661, 1180

\bibitem[\protect\citeauthoryear{{Orosz}, {McClintock}, {Aufdenberg},
  {Remillard}, {Reid}, {Narayan} \& {Gou}}{{Orosz}
  et~al.}{2011}]{2011ApJ...742...84O}
{Orosz} J.~A.,  {McClintock} J.~E.,  {Aufdenberg} J.~P.,  {Remillard} R.~A.,
  {Reid} M.~J.,  {Narayan} R.,    {Gou} L.,  2011, \apj, 742, 84

\bibitem[\protect\citeauthoryear{{Parker}}{{Parker}}{1958}]{1958ApJ...128..677P}
{Parker} E.~N.,  1958, \apj, 128, 677

\bibitem[\protect\citeauthoryear{{Parkinson}, {Tournear}, {Bloom}, {Focke},
  {Reilly}, {Wood}, {Ray}, {Wolff} \& {Scargle}}{{Parkinson}
  et~al.}{2003}]{2003ApJ...595..333P}
{Parkinson} P.~M.~S.,  {Tournear} D.~M.,  {Bloom} E.~D.,  {Focke} W.~B.,
  {Reilly} K.~T.,  {Wood} K.~S.,  {Ray} P.~S.,  {Wolff} M.~T.,    {Scargle}
  J.~D.,  2003, \apj, 595, 333

\bibitem[\protect\citeauthoryear{{Penev}, {Sasselov}, {Robinson} \&
  {Demarque}}{{Penev} et~al.}{2007}]{2007ApJ...655.1166P}
{Penev} K.,  {Sasselov} D.,  {Robinson} F.,    {Demarque} P.,  2007, \apj, 655,
  1166

\bibitem[\protect\citeauthoryear{{Pols}, {Schr{\"o}der}, {Hurley}, {Tout} \&
  {Eggleton}}{{Pols} et~al.}{1998}]{1998MNRAS.298..525P}
{Pols} O.~R.,  {Schr{\"o}der} K.-P.,  {Hurley} J.~R.,  {Tout} C.~A.,
  {Eggleton} P.~P.,  1998, \mnras, 298, 525

\bibitem[\protect\citeauthoryear{{Portegies Zwart}, {McMillan} \& et
  al.}{{Portegies Zwart} et~al.}{2009}]{2009NewA...14..369P}
{Portegies Zwart} S.,  {McMillan} S.,    et al. H.,  2009, \na, 14, 369

\bibitem[\protect\citeauthoryear{{Portegies Zwart} \& {Verbunt}}{{Portegies
  Zwart} \& {Verbunt}}{1996}]{1996A&A...309..179P}
{Portegies Zwart} S.~F.,  {Verbunt} F.,  1996, \aap, 309, 179

\bibitem[\protect\citeauthoryear{{Pylyser} \& {Savonije}}{{Pylyser} \&
  {Savonije}}{1988}]{1988A&A...191...57P}
{Pylyser} E.,  {Savonije} G.~J.,  1988, \aap, 191, 57

\bibitem[\protect\citeauthoryear{{Rasio}, {Tout}, {Lubow} \& {Livio}}{{Rasio}
  et~al.}{1996}]{1996ApJ...470.1187R}
{Rasio} F.~A.,  {Tout} C.~A.,  {Lubow} S.~H.,    {Livio} M.,  1996, \apj, 470,
  1187

\bibitem[\protect\citeauthoryear{{Skumanich}}{{Skumanich}}{1972}]{1972ApJ...171..565S}
{Skumanich} A.,  1972, \apj, 171, 565

\bibitem[\protect\citeauthoryear{{Tout}, {Pols}, {Eggleton} \& {Han}}{{Tout}
  et~al.}{1996}]{1996MNRAS.281..257T}
{Tout} C.~A.,  {Pols} O.~R.,  {Eggleton} P.~P.,    {Han} Z.,  1996, \mnras,
  281, 257

\bibitem[\protect\citeauthoryear{{Val Baker}, {Norton} \& {Quaintrell}}{{Val
  Baker} et~al.}{2005}]{2005A&A...441..685V}
{Val Baker} A.~K.~F.,  {Norton} A.~J.,    {Quaintrell} H.,  2005, \aap, 441,
  685

\bibitem[\protect\citeauthoryear{{Valsecchi} \& {Rasio}}{{Valsecchi} \&
  {Rasio}}{2014}]{2014ApJ...786..102V}
{Valsecchi} F.,  {Rasio} F.~A.,  2014, \apj, 786, 102

\bibitem[\protect\citeauthoryear{{van der Meer}, {Kaper}, {van Kerkwijk} \&
  {van den Heuvel}}{{van der Meer} et~al.}{2005}]{2005AIPC..797..623V}
{van der Meer} A.,  {Kaper} L.,  {van Kerkwijk} M.~H.,    {van den Heuvel}
  E.~P.~J.,  2005, in {Burderi} L.,  {Antonelli} L.~A.,  {D'Antona} F.,  {di
  Salvo} T.,  {Israel} G.~L.,  {Piersanti} L.,  {Tornamb{\`e}} A.,
  {Straniero} O.,  eds, Interacting Binaries: Accretion, Evolution, and
  Outcomes Vol.~797 of American Institute of Physics Conference Series, {On the
  mass distribution of neutron stars in HMXBs}.
pp 623--626

\bibitem[\protect\citeauthoryear{{van Kerkwijk}, {van Paradijs} \&
  {Zuiderwijk}}{{van Kerkwijk} et~al.}{1995}]{1995A&A...303..497V}
{van Kerkwijk} M.~H.,  {van Paradijs} J.,    {Zuiderwijk} E.~J.,  1995, \aap,
  303, 497

\bibitem[\protect\citeauthoryear{{Verbunt} \& {Phinney}}{{Verbunt} \&
  {Phinney}}{1995}]{1995A&A...296..709V}
{Verbunt} F.,  {Phinney} E.~S.,  1995, \aap, 296, 709

\bibitem[\protect\citeauthoryear{{Verbunt} \& {Zwaan}}{{Verbunt} \&
  {Zwaan}}{1981}]{1981A&A...100L...7V}
{Verbunt} F.,  {Zwaan} C.,  1981, \aap, 100, L7

\bibitem[\protect\citeauthoryear{{Weber} \& {Davis} Jr.}{{Weber} \&
  {Davis}}{1967}]{1967ApJ...148..217W}
{Weber} E.~J.,  {Davis} Jr. L.,  1967, \apj, 148, 217

\bibitem[\protect\citeauthoryear{{Wong}, {Valsecchi}, {Fragos} \&
  {Kalogera}}{{Wong} et~al.}{2012}]{2012ApJ...747..111W}
{Wong} T.-W.,  {Valsecchi} F.,  {Fragos} T.,    {Kalogera} V.,  2012, \apj,
  747, 111

\bibitem[\protect\citeauthoryear{{Yungelson}, {Lasota}, {Nelemans}, {Dubus},
  {van den Heuvel}, {Dewi} \& {Portegies Zwart}}{{Yungelson}
  et~al.}{2006}]{2006A&A...454..559Y}
{Yungelson} L.~R.,  {Lasota} J.-P.,  {Nelemans} G.,  {Dubus} G.,  {van den
  Heuvel} E.~P.~J.,  {Dewi} J.,    {Portegies Zwart} S.,  2006, \aap, 454, 559

\bibitem[\protect\citeauthoryear{{Zahn}}{{Zahn}}{1966}]{1966AnAp...29..489Z}
{Zahn} J.~P.,  1966, Annales d'Astrophysique, 29, 489

\bibitem[\protect\citeauthoryear{{Zahn}}{{Zahn}}{1975}]{1975A&A....41..329Z}
{Zahn} J.-P.,  1975, \aap, 41, 329

\bibitem[\protect\citeauthoryear{{Zahn}}{{Zahn}}{1977}]{1977A&A....57..383Z}
{Zahn} J.-P.,  1977, \aap, 57, 383

\bibitem[\protect\citeauthoryear{{Zahn}}{{Zahn}}{1989}]{1989A&A...220..112Z}
{Zahn} J.-P.,  1989, \aap, 220, 112

\bibitem[\protect\citeauthoryear{{Zahn}}{{Zahn}}{2008}]{2008EAS....29...67Z}
{Zahn} J.-P.,  2008, in {Goupil} M.-J.,  {Zahn} J.-P.,  eds, EAS Publications
  Series Vol.~29 of EAS Publications Series, {Tidal dissipation in binary
  systems}.
pp 67--90

\end{thebibliography}
\end{document}